\documentclass{article}
\usepackage{subcaption}
\usepackage{microtype}
\usepackage{graphicx}
\usepackage{booktabs} 
\usepackage{enumitem}
\usepackage{hyperref}
\usepackage{CJKutf8}
\usepackage{afterpage}
\usepackage{array}
\usepackage{tabularx}
\usepackage{makecell}
\usepackage{booktabs}
\usepackage{graphicx}

\usepackage[accepted]{icml2025}

\usepackage{amsmath}
\usepackage{amssymb}
\usepackage{mathtools}
\usepackage{amsthm}
\usepackage{tabularx}
\usepackage{multirow}
\usepackage{lineno}
\usepackage{graphicx} 
\usepackage{hyperref}
\usepackage[capitalize,noabbrev]{cleveref}

\theoremstyle{plain}

\theoremstyle{definition}

\theoremstyle{remark}

\usepackage[textsize=tiny]{todonotes}
\usepackage[T1]{fontenc}

\icmltitlerunning{Llasa: Scaling Train-Time and Inference-Time Compute for Llama-based Speech Synthesis}

\begin{document}

\twocolumn[
\icmltitle{Llasa: Scaling Train-Time and Inference-Time Compute for   \\
           Llama-based Speech Synthesis}




\begin{icmlauthorlist}
\icmlauthor{Zhen Ye}{ust}
\icmlauthor{Xinfa Zhu}{nwpu}
\icmlauthor{Chi-Min Chan}{ust}
\icmlauthor{Xinsheng Wang}{ust}
\icmlauthor{Xu Tan}{indep}
\icmlauthor{Jiahe Lei}{USTB}
\icmlauthor{Yi Peng}{ust}
\icmlauthor{Haohe Liu}{Surrey}
\icmlauthor{Yizhu Jin}{ust}
\icmlauthor{Zheqi Dai}{cuhk}
\icmlauthor{Hongzhan Lin}{hkbu}
\icmlauthor{Jianyi Chen}{ust}
\icmlauthor{Xingjian Du}{Rochester}
\icmlauthor{Liumeng Xue}{ust}
\icmlauthor{Yunlin Chen}{Mobvoi}
\icmlauthor{Zhifei Li}{Mobvoi}
\icmlauthor{Lei Xie}{nwpu}
\icmlauthor{Qiuqiang Kong}{cuhk}
\icmlauthor{Yike Guo}{ust}
\icmlauthor{Wei Xue}{ust}
\end{icmlauthorlist}

\icmlaffiliation{ust}{The Hong Kong University of Science and Technology}
\icmlaffiliation{nwpu}{ASLP Lab, Northwestern Polytechnical University}
\icmlaffiliation{USTB}{University of Science and Technology Beijing}
\icmlaffiliation{hkbu}{Hong Kong Baptist University}
\icmlaffiliation{cuhk}{Chinese University of Hong Kong}
\icmlaffiliation{Mobvoi}{Shanghai Mobvoi Information Technology Co., Ltd}
\icmlaffiliation{Rochester}{University of Rochester}
\icmlaffiliation{indep}{Independent Researcher}
\icmlaffiliation{Surrey}{University of Surrey}
\icmlcorrespondingauthor{Wei Xue}{weixue@ust.hk}


\vskip 0.3in
]



\printAffiliationsAndNotice{\icmlEqualContribution} 

\begin{abstract}
 
Recent advances in text-based large language models (LLMs), particularly in the GPT series and the o1 model, have demonstrated the effectiveness of scaling both training-time and inference-time compute. However, current state-of-the-art TTS systems leveraging LLMs are often multi-stage, requiring separate models (e.g., diffusion models after LLM), complicating the decision of whether to scale a particular model during training or testing. This work makes the following contributions: First, 
we explore the scaling of train-time and inference-time compute for speech synthesis. 
Second, we propose a simple framework Llasa for speech synthesis that employs a single-layer vector quantizer (VQ) codec and a single Transformer architecture to fully align with standard LLMs such as Llama.
Our experiments reveal that scaling train-time compute for Llasa consistently improves the naturalness of synthesized speech and enables the generation of more complex and accurate prosody patterns.  
Furthermore, from the perspective of scaling inference-time compute, we employ speech understanding models as verifiers during the search, finding that scaling inference-time compute shifts the sampling modes toward the preferences of specific verifiers, thereby improving emotional expressiveness, timbre consistency, and content accuracy.
In addition, we released the checkpoint and training code for our TTS model (1B, 3B, 8B) and codec model publicly available.  

\begin{itemize}
  \item \textbf{Models}: \href{https://huggingface.co/collections/HKUSTAudio/llasa-679b87dbd06ac556cc0e0f44}{Hugging Face Collection}
  \item \textbf{Llasa Training Code}: \href{https://github.com/zhenye234/LLaSA_training}{GitHub Repository}
  \item \textbf{Codec Training Code}: \href{https://github.com/zhenye234/X-Codec-2.0}{GitHub Repository}
  \item \textbf{Inference-time Scaling Code}: \href{https://github.com/zhenye234/LLaSA_inference}{GitHub Repository}
  \item \textbf{Demo page}: \href{https://llasatts.github.io/llasatts/}{Demo page}
  
\end{itemize}

\end{abstract}
 
\section{Introduction}

Recent years have witnessed the remarkable success of large language models (LLMs) in the text domain, represented by the GPT series\cite{brown2020language,achiam2023gpt,radford2019language}. These advances have demonstrated that increasing model size and training data consistently yields better performance across a wide array of natural language understanding and generation tasks. However, as the text domain approaches data saturation, new directions are emerging, such as “o1” models \cite{jaech2024openai} that emphasize extensive computational effort at test time—thereby exhibiting an inference-time scaling effect. By investing more resources during inference, these models are able to produce higher-quality outputs and handle more complex tasks, offering a flexible avenue for performance improvement after the training phase.

Meanwhile, text-to-speech (TTS) research has also made impressive strides. Many existing TTS systems focus on devising better model architectures—leveraging well-designed modules, larger datasets, and increased model size—to push synthesis quality ever higher. While these efforts have propelled the field forward, they also tend to narrow the community’s perspective: the focus on better architectures can overshadow investigations into broader, potentially transformative research questions. In contrast, the text LLM community has converged on a relatively standard framework—a simple Transformer model with a tokenizer—which allows researchers to concentrate on fundamental issues such as training-time scaling laws \cite{kaplan2020scaling}, inference-time scaling behaviors \cite{snell2024scaling}, and downstream adaptations (e.g., fine-tuning \cite{hu2021lora}, pruning, and quantization\cite{zhu2024survey}). Such a common design philosophy has catalyzed rapid progress and deeper insights into the text domain.

Motivated by these observations, we propose aligning TTS with the minimalist yet powerful paradigm of text LLMs. We introduce a single Transformer-based TTS model that relies on a well-designed speech tokenizer. More specifically,  our TTS system, named Llasa, is initialized from the Llama \cite{touvron2023Llama} model with an expanded vocabulary that incorporates speech tokens and is trained using
the next-token prediction paradigm.
Although this model may not always match the performance of highly customized TTS systems, its streamlined design creates a unified foundation for exploring a wider range of research questions—beyond architecture exploration.

In particular, we systematically investigate both training-time and inference-time scaling effects under this unified TTS framework. Our experiments show that scaling training-time compute (e.g., increasing model size or training data) not only improves naturalness but also enhances expressive prosody, effectively capturing the meaning conveyed in text
without explicit labels. Additionally, we examine the utility of inference-time scaling by incorporating speech understanding models as verifiers in the search process. We find that spending more computational effort during inference aligns generation outputs more closely with specific verifier biases, yielding better emotional expressiveness, timbre consistency, and content accuracy. Evaluations on LibriSpeech test sets \cite{panayotov2015librispeech}, seed-tts-eval \cite{anastassiou2024seed} and ESD datasets \cite{zhou2021emotional} demonstrate state-of-the-art results, and further highlight how in-context learning capabilities can be combined with search-based refinements to control factors such as speaker identity or emotion.

In summary, our paper makes several key contributions. We design a TTS model named Llasa that is fully aligned with standard LLM architectures by utilizing a single Transformer and a well-designed speech tokenizer, thereby creating a simple, flexible, and scalable system. Additionally, we find that increasing training-time compute for Llasa leads to significant improvements in speech naturalness and prosody accuracy, which reflects a deeper semantic understanding of the input text. We further demonstrate that scaling inference-time compute—achieved by incorporating speech understanding verifiers—enhances emotional expressiveness, timbre consistency, and content accuracy in synthesized speech. Furthermore, by providing open access to our models and frameworks, we aim to foster further research and development in the TTS community.

\section{Methods}
 
\subsection{Overview}
  
Our TTS framework is designed to fully align with the standard text LLM paradigm, keeping only two main components: (1) a tokenizer and (2) a single Transformer-based LLM. We initialize the Transformer parameters \(\phi\) from an existing LLM (e.g., Llama), and inherit its tokenizer for the text portion. Hence, the core new challenge is to convert raw speech waveforms into sequences of discrete tokens such that the Transformer can model them in an autoregressive manner. 

To achieve this, we introduce our speech tokenizer, X-codec2, which encodes waveforms into speech tokens and can decode them back to high-quality audio. Unlike some prior tokenizers for TTS, ours requires no additional information during decoding, ensuring that all aspects of the speech signal, such as content, prosody, and timbre, are captured by the LLM.

Formally, let:
1. \( \texttt{Tokenize}_\text{text}(X) = \{x_1, \dots, x_T\} \) be the text tokenizer, which converts input text \(X\) into \(T\) text tokens.
2. \( \texttt{Tokenize}_\text{speech}(Y) = \{y_1, \dots, y_S\} \) be the speech tokenizer, which converts a speech waveform \(Y\) into \(S\) discrete tokens.
3. \( \texttt{Detokenize}_\text{speech}(\{y_1, \dots, y_S\}) = \hat{Y} \) be the speech decoder, which reconstructs the waveform \(\hat{Y}\) from its token representation.

Given a dataset \(\mathcal{D} = \{(X_i, Y_i)\}_{i=1}^N\), where \(X_i\) is the text transcription and \(Y_i\) is the corresponding audio, we represent each pair \((X_i, Y_i)\) as a token sequence \((x_1, \dots, x_T, y_1, \dots, y_S)\). Our Transformer, with parameters \(\phi\), learns the joint distribution of text and speech tokens via 
\begin{align}
    &P(x_1, \ldots, x_T, y_1, \ldots, y_S)  \notag \\
  = & P(x_1, \ldots, x_T) \cdot  P(y_1, \ldots, y_S | x_1, \ldots, x_T) \notag
\end{align}

Since the text tokens \( \{x_1, \ldots, x_T\} \) are always given as input during training and inference, the model focuses on learning the conditional probability:
\begin{align}
    &P(y_1, \ldots, y_S | x_1, \ldots, x_T) \notag \\
    =& \prod_{s=1}^S P(y_s | x_1, \ldots, x_T, y_1, \ldots, y_{s-1}) \notag
\end{align}

Therefore, the loss is calculated over the speech tokens \( \{y_1, \ldots, y_S\} \). The objective is to minimize the negative log-likelihood:
\[
\mathcal{L} = - \sum_{s=1}^S \log P(y_s | x_1, \ldots, x_T, y_1, \ldots, y_{s-1})
\]
which makes the model learn to predict each speech token \( y_s \) conditioned on both the text tokens \( \{x_1, \ldots, x_T\} \) and the previously generated speech tokens \( \{y_1, \ldots, y_{s-1}\} \).

\subsection{Speech Tokenizer}
 
As highlighted by AudioLM \cite{borsos2023audiolm}, discrete speech representations can be categorized into both semantic tokens and acoustic tokens. Language modeling on semantic tokens excels at capturing high-level information such as content and emotion, while modeling with acoustic tokens focuses on low-level details, including timbre and other acoustic characteristics. Our X-codec2 tokenizer builds on the concepts from prior work X-codec \cite{ye2024codec}. We fuse these semantic and acoustic features into a unified codebook but introduce a crucial modification: replacing residual vector quantization with a single vector quantizer to ensure 1D causal dependency. This design naturally aligns with the left-to-right autoregressive mechanism of LLMs and also reflects the inherently left-to-right temporal structure of audio signals.
 
Our X-codec2 consists of three main components: the Encoder, the VQ module, and the Decoder.
 
\paragraph{Encoder}  
Given a raw speech waveform \(\mathbf{Y}\), we employ two separate encoders to derive its semantic and acoustic representations:
\begin{itemize}[leftmargin=1.5em]
    \item \textbf{Semantic Encoder} \(\mathrm{Enc}_s\): We adopt a pre-trained Wav2Vec2-BERT \cite{barrault2023seamless} to obtain multilingual speech semantic features that capture content and emotional cues.
    \item \textbf{Acoustic Encoder} \(\mathrm{Enc}_a\): Following the design in \citet{xin2024bigcodec} and \citet{kumar2024high}, this module uses multiple residual convolutional blocks with Snake activation functions to encode low-level acoustic details.
\end{itemize}
We then concatenate the two outputs to form a fused feature embedding,
\[
\mathbf{H} = \bigl[\mathrm{Enc}_s(\mathbf{Y}), \; \mathrm{Enc}_a(\mathbf{Y})\bigr],
\]
which serves as the input to our vector quantizer.
 
\paragraph{Vector Quantization}  
To obtain discrete tokens, we apply \(\mathrm{FSQ}(\cdot)\) \cite{mentzer2024finite} to \(\mathbf{H}\):
\[
\mathbf{H}_q \;=\; \mathrm{FSQ}(\mathbf{H}),
\]
where \(\mathbf{H}_q\) is the quantized feature. We adopt FSQ due to its stability in training and high codebook usage efficiency. Notably, FSQ does not require an explicit VQ objective term (e.g., codebook commitment loss), simplifying optimization during training.
 
\paragraph{Decoder}  
From the quantized representation \(\mathbf{H}_q\), we aim to reconstruct both \emph{semantic} and \emph{acoustic} information:
\begin{itemize}[leftmargin=1.5em]
    \item \textbf{Semantic Reconstruction}: We follow \citet{ye2024codec} and employ a semantic decoder to predict semantic features, using an \(\ell_2\) loss for reconstruction. It is worth noting that during inference, predicting semantic features is unnecessary; this component is designed to provide a supervisory signal to ensure the codebook retains sufficient semantic information.
    \item \textbf{Acoustic Reconstruction}: Following Vocos \cite{siuzdakvocos}, we replace the ConvNeXt backbone with a Transformer-based decoder that predicts the short-time Fourier transform (STFT) magnitude and phase; an inverse STFT (iSTFT) head then converts the predicted spectra back to time-domain waveforms.
\end{itemize}
 
\paragraph{Codec Training}  
The training process closely follows that of X-Codec, simultaneously optimizing both semantic and acoustic reconstruction. We incorporate a multi-period discriminator (MPD) \cite{kong2020hifi}, a multi-scale STFT (MS-STFT) discriminator, and a spectral discriminator, with FFT sizes \(\{78, 126, 206, 334, 542, 876, 1418, 2296\}\) \cite{anonymous2025scaling}, for adversarial training. Additionally, following \cite{anonymous2025scaling}, we incorporate a perceptual loss during the final steps of the training process to further enhance intelligibility.

\subsection{Scaling Train-time Compute}
Our primary goal in this section is not to locate a compute-optimal configuration for TTS, but rather to show that, akin to text-based LLMs, increasing train-time resources (either by enlarging the model or expanding the training dataset) consistently improves performance. Specifically, we investigate two scaling strategies:

\begin{itemize}[leftmargin=1.5em]  
    \item \textbf{Fix Training Data, Vary Model Size.} We fix the training data at \(250\text{k}\) hours and scale the size of the Transformer \(\phi\). Concretely, we adopt Llama~3.2 with 1B and 3B parameters, as well as Llama~3.1 with 8B parameters, which we denote as Llasa-1B-250k, Llasa-3B-250k, and Llasa-8B-250k, respectively, to observe how increased model capacity influences TTS quality.
    \item \textbf{Fix Model Size, Vary Training Data.} We choose an LLM initialized from Llama~3.2 1B and train on three progressively larger subsets of our dataset \(\mathcal{D}\), containing \(80\text{k}\), \(160\text{k}\), and \(250\text{k}\) hours of speech, respectively. Notably, the 80k and 160k subsets are randomly sampled as 1/3 and 2/3 partitions of the full 250k dataset, which we denote as Llasa-1B-80k, Llasa-1B-160k, and Llasa-1B-250k (identical to the previously mentioned Llasa-1B-250k model).
\end{itemize}

We evaluate these on two aspects:

\paragraph{Text Understanding Ability}  
A longstanding challenge in Text-to-Speech (TTS) technology is that TTS systems often fail to fully comprehend the meaning of text as humans do, which leads to mechanical pronunciation, lack of emotion, unnatural pauses, and difficulties in distinguishing homographs. Following BASE TTS \cite{lajszczak2024base}, we use seven categories of texts—Questions, Emotions, Compound Nouns, Complex Syntax, Foreign Words, Paralinguistics, and Punctuation—for English evaluation. Additionally, we propose seven categories tailored for Chinese: Questions, Emotions, Paralinguistics, Chinese Poetry, Rare Characters, Polyphonic Words, and Tongue Twisters (details are provided in Appendix \ref{testset}). In each case, the TTS system must exhibit a deeper textual understanding to produce natural and context-appropriate speech (e.g., correct pronunciation for polyphonic words and more expressive speech for emotional content). By examining the synthesized audio for each category, we measure how increased training data or parameter count benefits the system's text understanding ability.

\paragraph{In-context Learning Ability}  
We also evaluate the model’s zero-shot TTS capabilities \cite{wang2023neural}—whether it can produce intelligible, high-quality speech for speakers unseen during training. This aligns with prior zero-shot TTS protocols, which typically assess how well a model generalizes to new speaker identities, timbres, and emotional expressions with no additional fine-tuning.


 
\subsection{Scaling Inference-time Compute}

Recent research has begun exploring the scaling behavior of LLMs during inference, showing that additional computational resources—often in the form of sophisticated search strategies—can further enhance performance. Concretely, such approaches adjust the model’s output distribution at test time by generating multiple candidate outputs from a baseline model and then applying post-hoc filtering and refinement via verifiers or scoring mechanisms, thereby elevating the quality of the generated content. When extending this concept to text-to-speech (TTS), we hypothesize that generating multiple speech candidates and performing a targeted search among them can yield outputs that more closely match the task requirements. In line with prior work \cite{snell2024scaling,ma2025inference}, our search framework centers on two fundamental design choices: 

\paragraph{Verifier Selection}  
For TTS, many off-the-shelf speech understanding models can serve as verifiers (or reward models) to evaluate synthesized audio from multiple perspectives. These include speaker verification models for measuring timbre similarity, emotion representation models \cite{ma2023emotion2vec} for gauging emotional content, prosodic analyzers (e.g., pitch and duration \cite{li2023diverse,ren2022fastspeech2fasthighquality}) to ensure natural intonation and rhythm, speech quality metrics \cite{saeki2022utmos,reddy2021dnsmos} (e.g., SpeechMos) to evaluate naturalness, and automatic speech recognition (ASR) models \cite{radford2023robust,gao2023funasr} to assess transcription accuracy. By integrating feedback from these diverse verifiers, we ensure that the generated speech meets our requirements across multiple aspects, all in a fully automated process.

\paragraph{Algorithms}  
We categorize the two different reward methods as follows:  
\begin{itemize}[leftmargin=1.5em]
\item {\textbf{Output Reward Models (ORMs)}:}These models assess the speech segment only after it has been fully generated, evaluating it holistically. ORMs typically follow a simpler design but may be less efficient due to the absence of intermediate guidance signals. A common search strategy based on ORMs is the Best-of-N approach, where multiple candidate outputs are generated, scored using a reward model, and the highest-scoring output is selected.  
\item {\textbf{Process Reward Models (PRMs)}:} These models evaluate the generation process step by step, optimizing at each incremental stage (e.g., every second in a 10-second clip). Unlike conventional reward models that produce a single score for the final output, PRMs provide a sequence of scores, allowing for more fine-grained feedback throughout the generation process. While this approach enables detailed control, it also increases the risk of overfitting to intermediate rewards and of converging to local optima. The most common search algorithm leveraging PRMs is beam search, which systematically explores the solution space while optimizing both the sampling and evaluation of intermediate steps.  
\end{itemize}
In our experiments, we explore both PRMs and ORMs to analyze how different search strategies impact the final quality of synthesized speech. A more detailed discussion of these methods and their outcomes is provided in the next experiments section.

\begin{table*}[ht]
    \centering
    \caption{Comparison between different codec models. Bold values indicate the best for each token rate. We use token rate instead of bitrate because, from the perspective of LLMs, it is more intuitive: dividing the speech context window length by the token rate directly gives the generated audio duration in seconds.}
    \label{codec}
    \tiny
    \setlength{\tabcolsep}{5pt}
    \resizebox{\textwidth}{!}{
    \begin{tabular}{l|c|c|c|c|c|c|c|c|c|c}
    \toprule
    Model & 
    Token Rate & 
    \makecell{\scriptsize Codebook\\ \scriptsize Size} & 
    \makecell{\scriptsize Codebook\\ \scriptsize Layer} & 
    \makecell{\scriptsize Frame\\ \scriptsize Rate} & 
    \makecell{\scriptsize WER\\ $\downarrow$} &
    \makecell{\scriptsize STOI\\ $\uparrow$} &
    \makecell{\scriptsize PESQ-\\ \scriptsize WB$\uparrow$} &
    \makecell{\scriptsize PESQ-\\ \scriptsize NB$\uparrow$} &
    \makecell{\scriptsize SPK-\\ \scriptsize SIM$\uparrow$} &
    \makecell{\scriptsize UT\\ \scriptsize MOS$\uparrow$} \\
   \hline
    Ground Truth        
        & -      & -     & -    & -    & 1.96   & 1.00  & 4.64    & 4.55 & 1.00 & 4.09 \\ 
   \hline
    DAC 
        & 600   & 1024  & 12    & 50   & \textbf{2.00 }  & \textbf{0.95 } & \textbf{4.01  }  & \textbf{4.15 }&\textbf{ 0.95} & \textbf{4.00 }\\ 
    Encodec     
        & 600   & 1024  & 8    & 75   & 2.15   & 0.94  & 2.77    & 3.18 & 0.89 & 3.09 \\   \hline
    Encodec & 150  & 1024  & 2    & 75   & 4.90   & 0.85  & 1.56    & 1.94 & 0.60 & 1.58 \\
    DAC & 100   & 1024  & 2    & 50   & 13.27   & 0.73  & 1.13    & 1.40 & 0.32 & 1.29 \\
    SpeechTokenizer
        & 100   & 1024  & 2    & 50   & 3.92   & 0.77  & 1.25    & 1.59 & 0.36 & 2.28 \\
    Mimi    
        & 100   & 2048  & 8    & 12.5   & 2.96   & \textbf{0.91 } &  2.25   & 2.80 &\textbf{ 0.73 }& 3.56 \\
    X-codec       
        & 100   & 1024  & 2    & 50   &\textbf{ 2.49 }  & 0.86  &\textbf{ 2.33 }   &\textbf{ 2.88} & 0.72 & \textbf{4.21} \\  \hline
    BigCodec     
        & 80    & 8192  & 1    & 80   &\textbf{ 2.76}   & \textbf{0.93 } & \textbf{2.68}    & \textbf{3.27} & \textbf{0.84} &\textbf{ 4.11} \\
    WavTokenizer
        & 75    & 4096  & 1    & 75   & 3.98   & 0.90  & 2.13    & 2.63 & 0.65 & 3.79 \\
    Mimi      
        & 75    & 2048  & 6    & 12.5   & 3.61   & 0.89  & 1.99    & 2.51 & 0.65 & 3.38 \\
    Encodec   
        & 75    & 1024  & 1    & 75   & 28.92  & 0.77  & 1.23    & 1.48 & 0.25 & 1.25 \\ \hline
    DAC   
        & 50    & 1024  & 1    & 50   & 74.55  & 0.62  & 1.06    & 1.20 & 0.08 & 1.25 \\
    SpeechTokenizer
        & 50    & 1024  & 1    & 50   & 5.01   & 0.64  & 1.14    & 1.30 & 0.17 & 1.27 \\
    Mini
        & 50    & 2048  & 4    & 12.5   & 4.89   & 0.85  & 1.64    & 2.09 & 0.50 & 3.03 \\
    StableCodec 
        & 50    & 15625 & 2    & 25   & 5.12   & 0.91  & 2.24    & 2.91 & 0.62 & \textbf{4.23 }\\
    SemantiCodec
        & 50    & 32768/8192 & 2    & 25   & 6.89   & 0.84  & 1.66    & 2.18 & 0.58 & 2.71 \\
     
    X-codec        
        & 50    & 1024  & 1    & 50   & 3.42   & 0.83  & 1.84    & 2.38 & 0.52 & 4.05 \\  
    WavTokenizer        
        & 40    & 4096  & 1    & 40   & 11.20   & 0.85  & 1.62    & 2.06 & 0.48 &  3.57  \\
        \hline
    X-codec2 (ours) 
        & 50    & 65536 
        & 1 
        & 50 
        & \textbf{2.47} 
        &\textbf{ 0.92} 
        & \textbf{2.43} &
        \textbf{3.04} 
        & \textbf{0.82 }
        & 4.13 \\
 
    \bottomrule
    \end{tabular}
    }
\end{table*}

\section{Experiments}
In this section, first, we compare our proposed speech tokenizer with existing codecs to assess its effectiveness. Second, we evaluate the performance of our TTS systems. We explore the effect of scaling both train-time and inference-time compute and compare our models against other baseline TTS systems. Lastly, we evaluate the extensibility of our framework in Appendix \ref{asr}, particularly its applicability to speech understanding tasks, highlighting its versatility and potential for broader applications
\subsection{Codec Experiments}
\subsubsection{Training Details}
We train our codec model on a corpus of approximately 150k hours of multilingual speech, drawn from the Emilia dataset (En/Zh/De/Fr/Ja/Ko)\cite{he2024emilia} and MLS (En/Fr/De/Nl/Es/It/Pt/Pl)\cite{Pratap2020MLSAL}. All audio is sampled at 16\,kHz. We set the total downsampling ratio \(R\) to 320, use a codebook size of 65536, and employ a projection dimension of 8 in our VQ module. During training, we randomly crop 6-second segments from the audio. The learning rate is \(1\times10^{-4}\), preceded by a 3000-step warmup. In total, we train for 1.4 million steps, we activate perceptual loss at the final 0.2 million steps.
\subsubsection{Evaluation Results}
\label{codec_analyse}
For evaluation, we use the test-clean subset of LibriSpeech~\cite{panayotov2015librispeech}, which contains 2620 utterances at 16\,kHz.  

We evaluate our system using several metrics. a HuBERT-based ASR system for Word Error Rate (WER) \footnote{\url{https://huggingface.co/facebook/hubert-large-ls960-ft}}. Short-Time Objective Intelligibility (STOI). Perceptual Evaluation of Speech Quality (PESQ). A  WavLM-based speaker verification model for speaker similarity (SPK SIM) \footnote{\url{https://github.com/microsoft/UniSpeech/tree/main/downstreams/speaker_verification}}, and UTMOS \footnote{\url{https://github.com/tarepan/SpeechMOS}}.
 
We compare our codec against multiple baselines, including DAC \cite{kumar2024high}, SpeechTokenizer \cite{zhang2024speechtokenizer}, BigCodec \cite{xin2024bigcodec}, StableCodec \cite{parker2024scaling}, SemantiCodec \cite{liu2024semanticodec},   X-codec \cite{ye2024codec}, Mimi\cite{defossez2024moshi}, EnCodec \cite{defossez2022high}, and WavTokenizer \cite{ji2024wavtokenizer}. All baseline results are obtained using their official checkpoints. 

As shown in Table \ref{codec},  X-codec2 achieves the best performance at a token rate of 50 for most metrics. Moreover, its UTMOS score closely matches that of the ground truth, indicating that the reconstructed audio faithfully preserves the original speech quality. We also observe that certain models exceed the ground truth in UTMOS when operating at low token rates. We suspect this occurs because, under limited token constraints, the decoder behaves partly as a generative model—yielding plausible speech output but the alignment with the input was less precise. Additionally, we found that metrics such as WER at a low token rate can achieve good results by integrating speech semantic information, as demonstrated by models like Mini, X-codec, and SpeechTokenizer.  Another important observation is that the acoustic reconstruction capability of codecs at low token rates remains relatively limited. For instance, DAC operating at 600 tokens achieves a SPK SIM  of 0.95 and a PESQ score exceeding 4. In contrast, current codecs at lower token rates attain SPK SIM values below 0.85 and PESQ scores around 3. However, compared to earlier models like DAC and Encodec, there has been significant improvement in performance at low token rates. We believe that there is substantial potential for further advancements in low token rate codecs.

\begin{figure*}[t]
    \centering
    \includegraphics[width=0.99\textwidth]{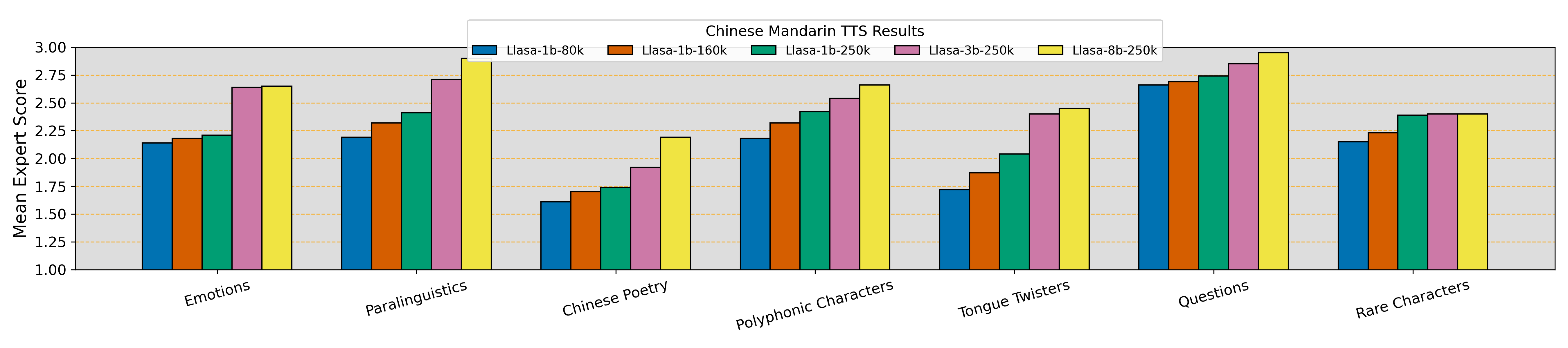} \\ 
    \includegraphics[width=0.99\textwidth]{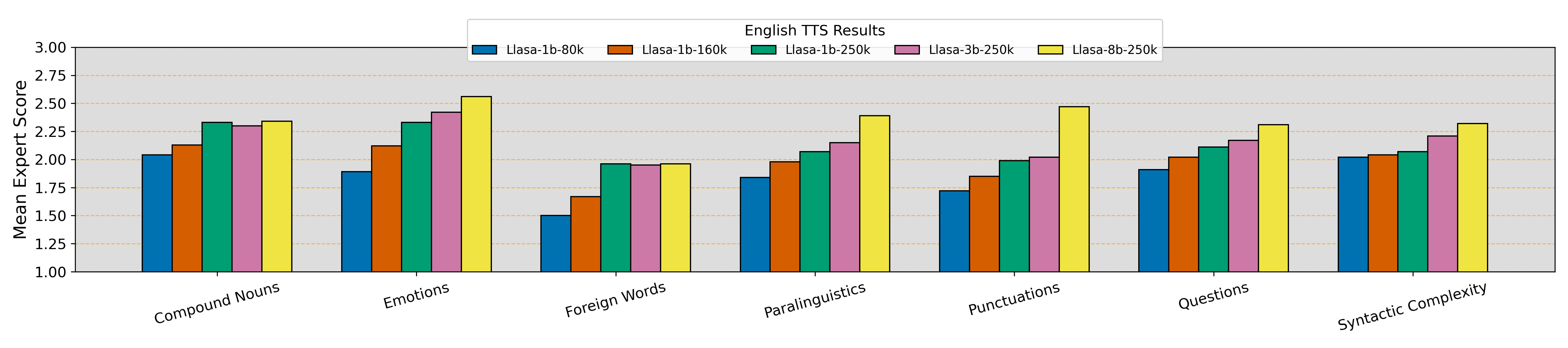} 
    \caption{Comparison of mean expert score for Chinese and English}
    \label{fig:mos_comparison}
\end{figure*}
 
\subsection{TTS experiments}
 
\subsubsection{Experimental details}
 
All TTS models are trained for 3 epochs with a batch size of 2 million tokens and a maximum learning rate of 5e-5. We employ a cosine learning rate schedule with a warmup phase covering 3\% of an epoch, and the final learning rate is set to 10\% of the peak learning rate. During training, text sequences are tokenized and placed on the left, followed by tokenized speech sequences on the right, forming a concatenated sequence that is then cropped to a maximum length of 2048 tokens.

Our training dataset integrates multiple high-quality speech datasets, including Libriheavy \cite{kang2024libriheavy}, the Chinese-English subset from the Emilia corpus \cite{he2024emilia}, WenetSpeech4TTS \cite{ma2024wenetspeech4tts}, and our internal data. This diverse collection constitutes a large-scale training corpus of 250,000 hours, comprising mixed Mandarin Chinese and English speech data. All textual content maintains its original punctuation.

Our TTS systems are evaluated through a series of comprehensive experiments designed to assess various aspects of performance. 

\textbf{Evaluation of Text understanding ability }
 
Similar to \cite{lajszczak2024base}, we evaluated the models' text understanding capabilities by focusing on their ability to accurately comprehend and synthesize speech from complex textual inputs. This assessment aimed to evaluate the text understanding abilities of TTS systems. The full testset is in Appendix \ref{testset}, each sentence was synthesized five times by each model. Due to the absence of speech prompts, timbre and style were generated randomly. The evaluation criteria are also in Appendix \ref{testset}; a linguistic expert rated the TTS outputs using a discrete 3-point scale, and we calculated the average score for each category.

\textbf{Evaluation of In-Context Learning Ability}
To assess the in-context learning capability of our model, we conducted experiments on three test sets: Seed-TTS-Eval, LibriSpeech test-clean, and the ESD dataset.  

The first two datasets primarily evaluate the model's ability to clone the voice of an unseen speaker given a short speech clip, focusing on speaker similarity. For both, speaker similarity (SIM) and Word Error Rate (WER) are used as key evaluation metrics:  

Seed-TTS-Eval\footnote{\url{https://github.com/BytedanceSpeech/seed-tts-eval}} \cite{anastassiou2024seed} consists of three subsets: test-zh, test-en, and test-hard. These experiments focus on cross-sentence speaker similarity and the generation of intelligible speech.  
LibriSpeech test-clean \cite{panayotov2015librispeech} is a widely used benchmark for English zero-shot TTS evaluation \cite{wang2023neural}, providing a standardized setting to assess the model’s ability to generate natural and intelligible speech from unseen speakers.   

The third test set, ESD (Emotional Speech Dataset)\cite{zhou2021seen,zhou2021emotional}, evaluates the model's ability to clone emotions in speech. This dataset consists of 10 native English speakers and 10 native Chinese speakers, covering five emotion categories: neutral, happy, angry, sad and surprised. For reproducibility, we selected the longest utterance from each speaker as the prompt and the second longest as the ground truth, resulting in a total of 100 evaluation samples (50 English, 50 Chinese). We used Emotion2Vec-Plus-Large \cite{ma2023emotion2vec} \footnote{\url{https://huggingface.co/emotion2vec/emotion2vec\_plus\_large}} to measure emotional similarity.  

Through these evaluations, we aimed to provide a comprehensive assessment of our TTS models, ensuring their effectiveness in speaker identity preservation, intelligibility, and emotional expressiveness across diverse linguistic and contextual challenges.
\subsubsection{Scaling Train-time Compute}

\textbf{Text Understanding Ability}
Figure \ref{fig:mos_comparison} presents expert scores for Chinese TTS tasks, including Emotion, Paralinguistics, Poetry, Polyphonic Characters, Tongue Twisters, Questions, and Rare Characters. Increasing both model size (1B → 3B → 8B) and training data (80k → 160k → 250k hours) generally improves performance. Specifically, simpler tasks like Questions already achieve strong results, with only marginal gains from further scaling. In contrast, larger models (e.g., 8B parameters) yield significant improvements in emotional speech, Chinese poetry, and tongue twisters, where deeper semantic comprehension is essential. Meanwhile, rare characters benefit most from broader data coverage, as increasing model size alone has little effect. We also evaluate our models on English TTS tasks—Compound Nouns, Emotions, Foreign Words, Paralinguistics, Punctuations, Questions, and Syntactic Complexity. As with Chinese, scaling up both the model size (1B → 3B → 8B) and training data (80k → 160k → 250k hours) generally yields higher expert scores.  We observe that both Compound Nouns and Foreign Words primarily benefit from increased training data rather than model scaling, suggesting that a wider variety of examples is necessary for correct pronunciation and lexical coverage.  

\textbf{In-context Learning Ability.  }
Based on Table \ref{seedttseval}, Table \ref{emosim}, and Table \ref{librispeecheval}, we observe that as the model size and training data increase, the metrics for speaker similarity, word error rate, and emotional similarity consistently improve, reflecting the enhancement of in-context learning ability.

\begin{figure*}[t]
    \centering
    \begin{subfigure}{0.495\textwidth}
        \centering
        \includegraphics[width=\textwidth]{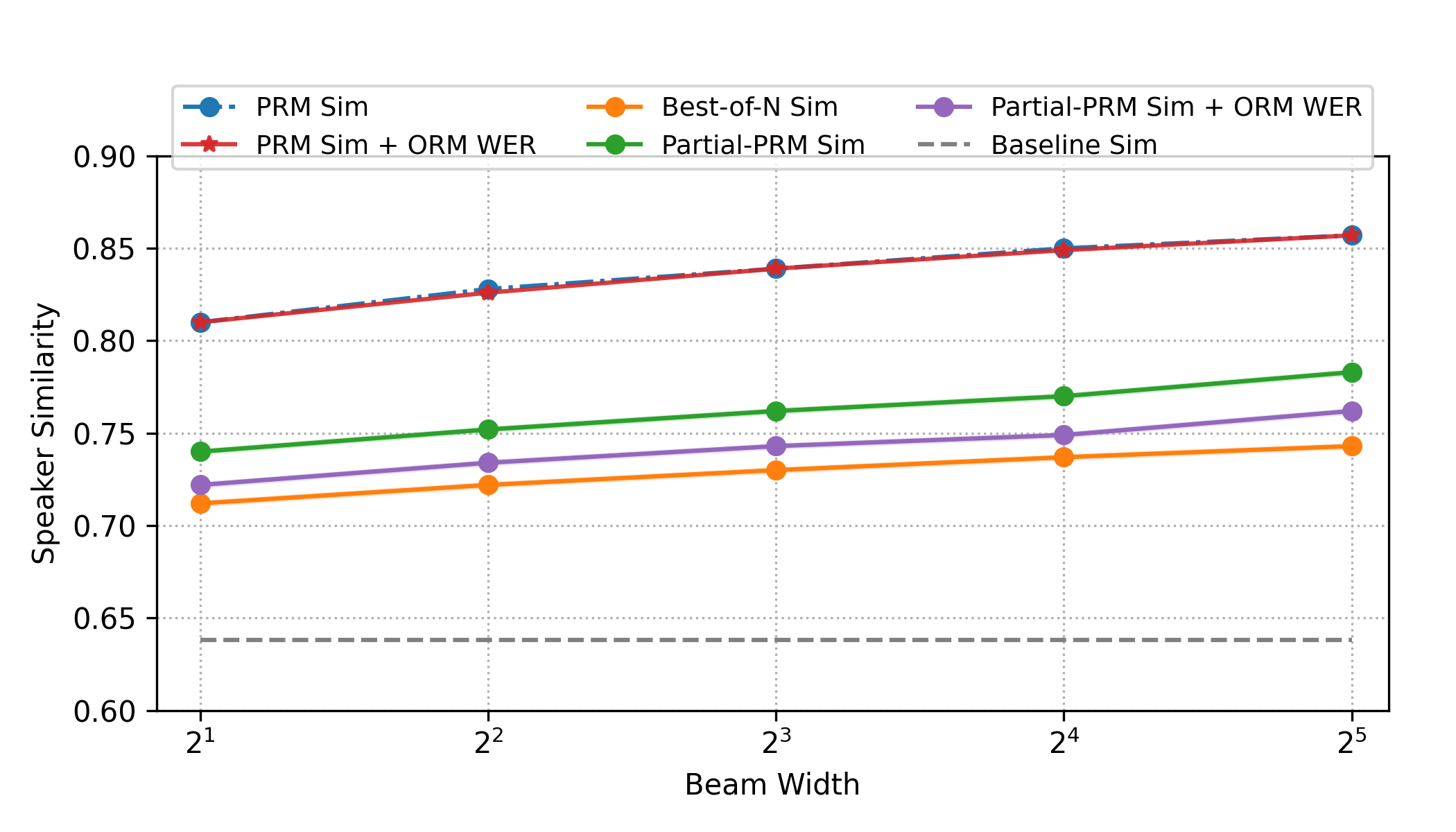}
        
    \end{subfigure}
    \hfill
    \begin{subfigure}{0.495\textwidth}
        \centering
        \includegraphics[width=\textwidth]{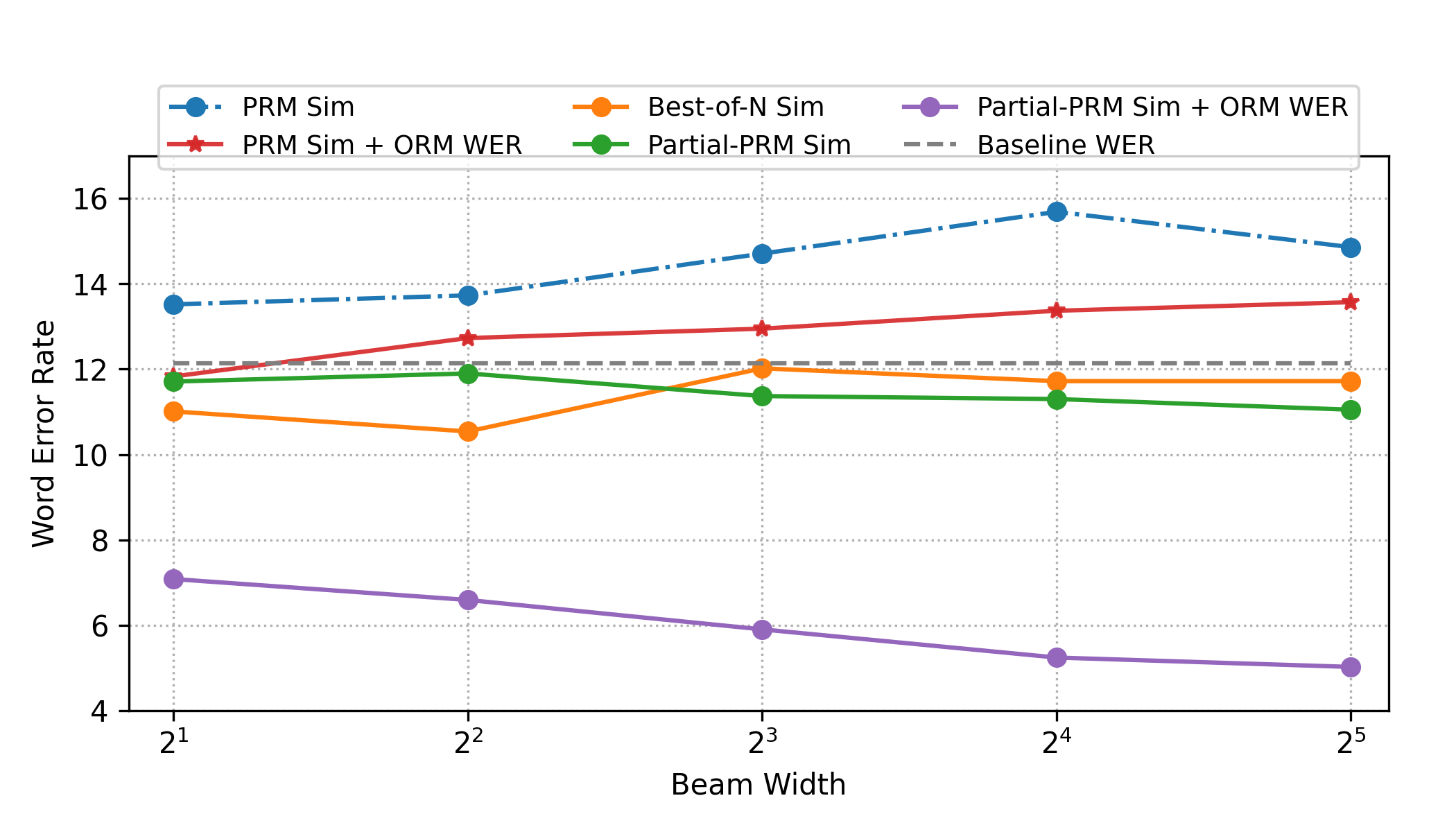}
  
    \end{subfigure}
    \caption{Illustration of inference-time compute scaling for speaker similarity and word error rates.} 
    \label{scaletest}
\end{figure*}

\subsubsection{Scaling inference-time Compute}
In this section, we conduct a series of experiments to explore how increasing inference-time compute affects the performance of different strategies, such as PRM and ORM. All experiments are conducted using the Llasa-1B-250K model and evaluated on seed-tts-eval test-hard testset.  We begin by evaluating zero-shot TTS performance using  SIM-O  and WER  metrics, with direct inference serving as the baseline. Our initial goal is to improve SIM, utilizing two key strategies: PRM with beam search and  ORM  with best-of-\(N\). The similarity verifier is a WavLM-finetuned speaker verification model.

For beam search, we generate \(B\) candidate beams conditioned on the speech prompt, where each beam is expanded by \(M\) tokens per step. We set \(M=25\), corresponding to \(0.5\)~seconds in our experiments. Each beam then expands into \(N=16\) new candidate sequences. From the resulting pool of \(B\times N\) sequences, we select the top~\(B\) based on their similarity scores. This process repeats until an end-of-sequence (EOS) token is generated or the sequence length reaches~\(2048\).

For best-of-\(N\), we generate multiple independent responses and select the one with the highest similarity score as the final output. To match the compute budget of beam search, we also produce \(B\times N\) candidates.

As shown in Figure~\ref{scaletest}, our simplest method for improving SIM is best-of-\(N\). The \textcolor{orange}{orange line} indicates that as inference-time compute grows, SIM improves markedly. Next, the \textcolor{blue}{blue line} shows that PRM beam search outperforms ORM under the same compute budget.

However, when we also aim to optimize WER ( using Whisper~Large~v3 as a verifier), we select the lowest-WER candidate at the final PRM step from the \(B\times N\) sequences. The \textcolor{red}{red line} reveals that WER remains poor, especially for larger beam widths, sometimes lagging behind the \textbf{baseline}. We suspect that maximizing SIM through PRM leads to locally optimal speech with inadequate diversity. To address this, we propose a partial PRM strategy, applying PRM only during the first \(n\)~seconds, then switching to ORM. Here, \(n=2\) in our experiments. This hybrid approach (the \textcolor{green}{green line}) achieves higher SIM than best-of-\(N\) while maintaining WER near ground truth, indicating sufficient diversity. Finally, substituting the later ORM step with a WER-based verifier (the \textcolor{purple}{purple line}) simultaneously boosts both SIM and WER as inference-time compute increases, demonstrating that this mixed strategy strikes an effective balance between speaker similarity and transcription accuracy.
 
From another perspective, we also compare the impact of inference-time scaling across different model sizes and test sets. In these evaluations, we fix the beam width at 16. As shown in Tables \ref{librispeecheval}, \ref{emosim}, and \ref{seedttseval}, our results show that inference-time scaling not only improves speaker similarity but also enhances emotion similarity.  
Additionally, we generally observe that larger models benefit more from inference-time scaling. However, for certain relatively simple tasks and metrics, the performance gap between large and small models after inference-time scaling remains minimal.  This finding suggests a possible parallel with text-based LLMs: rather than solely focusing on scaling model training, in some settings, a more efficient approach may be to train smaller models with less compute and then leverage inference-time compute to enhance outputs.

\begin{table*}[ht]
 
\centering
\caption{Results of Llasa and recent TTS models on the SEED test sets. † denotes close-sourced models. SIM-o is computed with the original speech and SIM-r with the reconstructed speech. For Llasa series, sim-o values include sim-r in parentheses. For WER, we employ Whisper-large-v3 \cite{radford2023robust} and Paraformerzh \cite{gao2023paraformerfastaccurateparallel} as the automatic speech recognition (ASR) engines for English and Mandarin, respectively. For SIM, we use WavLM-large fine-tuned on the speaker verification task \cite{chen2022wavlm}}
\small
\label{tab:results}
\begin{tabular}{l|cc|cc|cc}
\hline
\textbf{Model} & \multicolumn{2}{c|}{\textbf{test-zh}} & \multicolumn{2}{c|}{\textbf{test-en}} & \multicolumn{2}{c}{\textbf{test-hard}} \\
                & \textbf{CER ↓} & \textbf{sim-o ↑} & \textbf{WER ↓} & \textbf{sim-o ↑} & \textbf{WER ↓} & \textbf{sim-o ↑} \\ 
\hline
Human               & 1.26 & 0.755 & 2.14 & 0.734 & -    & -       \\
Our Codec Resyn.    & 1.92 & 0.677 & 2.91 & 0.619 & -    & -       \\ \hline
Seed-TTS              & 1.12 & 0.796 & 2.25 & 0.762 & 7.59 & 0.776  \\
FireRedTTS          & 1.51 & 0.635 & 3.82 & 0.460 & 17.45& 0.621  \\
MaskGCT             & 2.27 & 0.774 & 2.62 & 0.714 & 10.27& 0.748  \\
E2 TTS (32 NFE)  & 1.97 & 0.730 & 2.19 & 0.710 & -    & -       \\
F5-TTS (32 NFE)     & 1.56 & 0.741 & 1.83 & 0.6l47 & 8.67 & 0.713  \\  
CosyVoice           & 3.63 & 0.723 & 4.29 & 0.609 & 11.75& 0.709  \\
CosyVoice 2         & 1.45 & 0.748 & 2.57 & 0.652 & 6.83 & 0.724  \\ \hline

\multicolumn{7}{c}{\textbf{ Train-time Scaling}} \\ \hline
Llasa-1B-80k       & 2.69 & 0.648 (0.779) & 3.71 & 0.541 (0.685) & 17.11 & 0.618 (0.765) \\
Llasa-1B-160k      & 2.22 & 0.658 (0.783) & 3.60 & 0.563 (0.701) & 16.73 & 0.627 (0.770) \\
Llasa-1B-250k      & 1.89 & 0.669 (0.794) & 3.22 & 0.572 (0.708) & 12.13 & 0.638 (0.779) \\
Llasa-3B-250k      & 1.60 & 0.675 (0.792) & 3.14 & 0.579 (0.708) & 13.37 & 0.652 (0.782) \\
Llasa-8B-250k      & 1.59 & 0.684 (0.798) & 2.97 & 0.574 (0.706) & 11.09 & 0.660 (0.787) \\ \hline
\multicolumn{7}{c}{\textbf{Partial PRM (spk sim)}} \\ \hline
Llasa-1B-80k       & 1.52 & 0.811 (0.849) & 2.30 & 0.761 (0.798) & 16.09 & 0.759 (0.824) \\
Llasa-1B-160k      & 1.29 & 0.815 (0.851) & 2.29 & 0.774 (0.804) & 14.10 & 0.768 (0.830) \\
Llasa-1B-250k      & 1.11 & 0.818 (0.855) & 2.03 & 0.781 (0.809) & 11.30 & 0.773 (0.833) \\
Llasa-3B-250k      & 1.06 & 0.824 (0.856) & 1.89 & \textbf{0.784} (0.812) & 11.22 & 0.780 (0.836) \\
Llasa-8B-250k      & 1.04 & \textbf{0.827} (0.856) & 1.84 & 0.783 (0.806) & 10.59 & \textbf{0.785} (0.839) \\ \hline
\multicolumn{7}{c}{\textbf{Partial PRM (spk sim)+ORM (WER)}} \\ \hline
Llasa-1B-80k      & 0.53 & 0.809 (0.840) & 1.43 & 0.761 (0.792) & 7.22 & 0.732 (0.789) \\
Llasa-1B-160k     & 0.53 & 0.812 (0.841) & 1.49 & 0.775 (0.798) & 6.35 & 0.745 (0.799) \\
Llasa-1B-250k     & \textbf{0.45} & 0.818 (0.845) & 1.46 & 0.782 (0.801) & 5.24 & 0.750 (0.803) \\
Llasa-3B-250k     & 0.50 & 0.823 (0.848) & \textbf{1.31} & 0.783 (0.803) & 5.39 & 0.759 (0.808) \\
Llasa-8B-250k     & 0.47 & 0.825 (0.848) & 1.39 & 0.783 (0.799) & \textbf{4.38} & 0.767 (0.812) \\ \hline

Llasa-8B-250k        & \multicolumn{4}{c|}{{Chunking: if $\text{len(char)} > 100 \rightarrow 2$ chunks, $> 200 \rightarrow 3$ chunks, $\dots$}} &  \textbf{3.12} & 0.770 (0.791) \\ \hline

\end{tabular}
\label{seedttseval} 
 
\end{table*}

\begin{table}[ht]
\centering
\caption{Objective performance comparison on \textit{continuation} zero-shot speech synthesis tasks. WER-H (\%) denotes evaluation with the HuBERT-Large ASR model. SIM-o is computed with the original speech and SIM-r with the reconstructed speech.}
\begin{tabular}{l|ccc}
\hline
\multirow{2}{*}{System} & \multicolumn{3}{c}{Continuation} \\
& WER-H & SIM-o & SIM-r \\
\hline
Ground Truth & 2.15 & 0.668 & - \\
Our Codec  Resyn. & 2.49 &  0.580 & 0.638 \\
\hline
ELLA-V    & 2.91 & 0.303 & 0.340 \\
VALL-E R    & 2.32 & 0.363 & 0.397 \\
 
CLaM-TTS   & 2.36 & 0.477 & 0.513 \\
VALL-E   & 3.8 & - & 0.508 \\
VALL-E 2   & 2.32 & 0.504 & 0.529 \\
Voicebox   & 2.0 & 0.593 &  0.616 \\

MELLE & 1.98 & 0.508 & 0.539 \\
\hline
 
\multicolumn{3}{c}{\textbf{ Train-time Scaling}} \\ \hline

Llasa-1B-80k  & 2.57 & 0.457  & 0.614\\
Llasa-1B-160k & 2.48 & 0.475 & 0.625 \\
Llasa-1B-250k  & 2.47 & 0.478  & 0.627 \\
Llasa-3B-250k & 2.35  & 0.484 & 0.628 \\
Llasa-8B-250k &  2.29 & 0.483 & 0.626 \\
\hline
\multicolumn{3}{c}{\textbf{Partial PRM (spk sim)}} \\ \hline
Llasa-1B-80k   & 2.43 & 0.699  & 0.738 \\
Llasa-1B-160k   & 2.36 & 0.712  & 0.744 \\
Llasa-1B-250k   & 2.37 & 0.712  & 0.743 \\
Llasa-3B-250k & 2.26  & \textbf{0.715} &\textbf{ 0.745} \\
Llasa-8B-250k & 2.24  & 0.714 & 0.741 \\
\hline
\multicolumn{3}{c}{\textbf{Partial PRM (spk sim)+ORMs (WER)}} \\ \hline
Llasa-1B-80k   & 1.76 & 0.700   & 0.738 \\
Llasa-1B-160k   & 1.66 & 0.710  & 0.743 \\
Llasa-1B-250k   &1.62 & 0.712 & 0.744\\
Llasa-3B-250k  & 1.57 & 0.714 & 0.742\\
Llasa-8B-250k  &  \textbf{1.49} & 0.714 &  0.740 \\
\hline
\end{tabular}
 
\label{librispeecheval} 
\end{table}

\begin{table}[ht] 
    \centering
     \caption{ Evaluation results for Emotion Similarity using Emotion2Vec-Plus-Large.  }
    \begin{tabular}{lcc}
        \toprule
        Model & en & zh \\
        \midrule
         GT  & 0.94  & 0.94  \\
        \midrule
        \multicolumn{3}{c}{\textbf{Train-time scaling}} \\
        \midrule
        Llasa-1B-80k  & 0.753  & 0.815  \\
        Llasa-1B-160k & 0.762  & 0.822  \\
        Llasa-1B-250k & 0.768  & 0.836  \\
        Llasa-3B-250k & 0.769  & 0.852  \\
        Llasa-8B-250k & 0.778  & 0.861  \\
        \midrule
        \multicolumn{3}{c}{\textbf{Process Reward Models (emotion sim)}} \\
        \midrule
        Llasa-1B-80k  & 0.933  & 0.970 \\
        Llasa-1B-160k & 0.936  & 0.971 \\
        Llasa-1B-250k & 0.937  & 0.974 \\
        Llasa-3B-250k & 0.949  & 0.975 \\
        Llasa-8B-250k & 0.951  & 0.974 \\
        \bottomrule
    \end{tabular}
   
    \label{emosim}
\end{table}

\subsubsection{Compare with baseline model}
 
In the previous sections, we analyzed our codec and explored the impact of scaling train-time and inference-time compute on the performance of TTS systems. In this section, we directly compare our model against other TTS baselines.

For Seed-TTS-Eval, we select the following baseline models: Seed-TTS \cite{anastassiou2024seed}, MaskGCT \cite{wang2024maskgct}, E2-TTS (32 NFE)$^\dagger$ \cite{eskimez2024e2}, F5-TTS (32 NFE) \cite{chen2024f5}, CosyVoice \cite{du2024cosyvoice}, CosyVoice 2 \cite{du2024cosyvoice2}, and FireRedTTS \cite{guo2024fireredtts}. Our results are taken from the original papers whenever available; otherwise, they are sourced from CosyVoice 2.

The results, as shown in Table \ref{seedttseval}, indicate that for direct inference, our model achieves WER performance comparable to these state-of-the-art (SOTA) models. However, one notable limitation of our approach is SIM-O. As discussed earlier, the reconstruction capability of single-token codecs remains constrained compared to mel-based vocoder reconstructions or RVQ codec-based reconstructions used in the baselines.

For LibriSpeech test-clean, we compare against the following baselines: ELLA-V \cite{song2024ella}, VALL-E R \cite{han2024vall}, CLaM-TTS \cite{kimclam}, VALL-E \cite{wang2023neural}, VALL-E 2 \cite{chen2024vall}, Voicebox \cite{le2024voicebox}, and MELLE \cite{meng2024autoregressive}. 

As shown in Table \ref{librispeecheval}, we observe similar trends in continuous TTS for LibriSpeech test-clean. However, an interesting finding is that our model achieves a high SIM-r score, particularly on LibriSpeech test-clean, where our best SIM-R (0.626) is already very close to the ground truth (GT) codec resynthesis (0.638). Given that the continuous generation task is fully aligned with the autoregressive training paradigm, this suggests that, from a generative modeling perspective, a single Transformer-based architecture does not inherently suffer disadvantages from metric Sim-r and WER compared to carefully designed AR+NAR hybrid architectures. The only drawback of Sim-o may arise at the system level, particularly in the final step of codec acoustic reconstruction, where converting the intermediate representation back into a waveform may introduce limitations due to single VQ as mentioned in Sec \ref{codec_analyse}.

From an inference-time scaling perspective, our approach outperforms all baselines. While this comparison might not be entirely fair—as our inference process utilizes more computational resources—it presents an alternative viewpoint: if the goal is to achieve the best possible quality, disregarding computational constraints, inference-time scaling provides a viable solution. Notably, as shown in table \ref{seedttseval} our model achieves a WER of 3.12 on the test-hard of Seed-TTS-Eval, demonstrating that allocating more compute at inference time is particularly beneficial for synthesizing challenging speech, effectively addressing cases where previous models have struggled.

\section{Extending to Speech Understanding Tasks}
\label{asr}
While we have primarily demonstrated the viability of our single Transformer + tokenizer approach for TTS, we also explored its performance on speech understanding task, in particular, ASR. The only modification is to swap the position of speech and text tokens: speech tokens come first, followed by text tokens, and during training we apply the cross-entropy loss solely to the text tokens. We use the same tokenizer X-codec2. ASR models are trained on Libriheavy \cite{kang2024libriheavy}, MLS English \cite{pratap2020mls}, and GigaSpeech \cite{chen2021gigaspeech} under a learning rate of \(2\times10^{-5}\), a batch size of 1M tokens, a 0.03 warmup ratio, and 2 training epochs. Table~\ref{tab:asr_results} presents ASR results on LibriSpeech. Notably, on the test-clean set, our model is competitive with Whisper Large v3. Performance on test-other, however, remains weaker—likely due to our smaller, relatively clean training set and the lack of data augmentation. Despite these limitations, our experiments confirm that an entirely discrete ASR paradigm operating on quantized speech tokens can still achieve promising results, comparable in many respects to mainstream approaches like Whisper, which relies on continuous Mel features.

\begin{table}[ht]
    \centering
    \caption{ASR Performance on LibriSpeech Test Sets}
    
    \label{tab:asr_results}
    \begin{tabular}{lcc}
        \toprule
        \textbf{Model} & \textbf{Test Clean  } & \textbf{Test Other } \\
        \midrule
        whisper large v3 & 1.8 & 3.6 \\
        whisper large v2 & 2.7 & 5.2 \\ \hline
        Llasa-asr-1b & 2.3 & 7.2 \\
        Llasa-asr-3b & 1.9 & 5.9 \\
        \bottomrule
    \end{tabular}
\end{table}

\section{Related work}
 
\subsection{Scaling Train-time and Inference-time Compute}
 
In the text domain, large language models (LLMs) like GPT \cite{brown2020language,kaplan2020scaling} show that increasing data, compute, and model size enhances performance, pushing LLMs toward cognitive intelligence. However, further scaling during training faces limits due to data and compute constraints. To overcome this, a scaling law during testing is proposed: greater computational effort in inference improves performance \cite{ji2025test}. Techniques like repeat sampling, self-correction, and tree search enhance reasoning depth and accuracy in complex tasks. While this principle has been extensively validated in the text domain, its impact on speech remains largely unexplored. Base-TTS \cite{lajszczak2024base} provides a brief investigation into the emergent abilities arising from scaling both model size and data volume but does not separately examine the effects of data scaling versus model scaling. Additionally, it does not explore performance across different languages. To the best of our knowledge, we are the first to systematically investigate inference scaling in the speech modality.
 
\subsection{LLM-based TTS}
 
Significant progress has been made in using large language models (LLMs) for TTS tasks, including multi-speaker synthesis \cite{speartts} and zero-shot voice cloning \cite{wang2023neural,chen2024vall}. VALL-E \cite{wang2023neural} pioneered treating TTS as a conditional language modeling problem by converting waveforms into neural codec tokens. However, it employed a multi-stage approach—a coarse autoregressive (AR) model followed by a non-autoregressive (NAR) residual model—which complicates training and inference. Extensions like VALL-E X \cite{vallex} enable cross-lingual synthesis, and Spear-TTS \cite{speartts} integrates multiple AR models to support multi-speaker TTS with minimal supervision. Recent TTS systems have often combined an AR language model with additional components, such as diffusion \cite{tortoisetts,basetts,anastassiou2024seed,du2024cosyvoice,guo2024fireredtts}, to generate more natural, controllable speech when trained on large datasets. Although these multi-stage pipelines can yield high-quality results, they remain cumbersome and less amenable to large-scale training. On the other hand, single-stage systems like MELL-E \cite{melle} and KALL-E \cite{kalle} avoid multi-stage generation but rely on continuous acoustic features (e.g., spectrograms or latent variables). Storing and processing these features at scale can be prohibitive, hindering training on tens or hundreds of billions of tokens. In contrast, our approach uses a single-stage AR Transformer that directly models discrete speech tokens, similar to how text LLMs handle words and subwords. This design avoids multi-stage complexity and the large memory footprint of continuous representations, making it far more scalable while retaining the flexibility of a standard LLM.

\section{Conclusion}
    
This paper presents Llasa, a scalable TTS system that aligns with text LLM architectures, using a single Transformer and a tokenizer. We systematically explore train-time and inference-time compute scaling, showing that larger models and datasets improve speech naturalness, prosody, and text comprehension. Additionally, inference-time scaling, leveraging speech understanding models as verifiers, enhances speaker similarity, emotional expressiveness, and content accuracy. 
Our experiments confirm state-of-the-art performance with strong zero-shot TTS capabilities. 
We release our models publicly to drive further research. 

\nocite{langley00}
\bibliography{example_paper}

\begin{thebibliography}{65}
\providecommand{\natexlab}[1]{#1}
\providecommand{\url}[1]{\texttt{#1}}
\expandafter\ifx\csname urlstyle\endcsname\relax
  \providecommand{\doi}[1]{doi: #1}\else
  \providecommand{\doi}{doi: \begingroup \urlstyle{rm}\Url}\fi

\bibitem[zho(2022)]{zhou2021emotional}
Emotional voice conversion: Theory, databases and esd.
\newblock \emph{Speech Communication}, 137:\penalty0 1--18, 2022.
\newblock ISSN 0167-6393.

\bibitem[Achiam et~al.(2023)Achiam, Adler, Agarwal, Ahmad, Akkaya, Aleman, Almeida, Altenschmidt, Altman, Anadkat, et~al.]{achiam2023gpt}
Achiam, J., Adler, S., Agarwal, S., Ahmad, L., Akkaya, I., Aleman, F.~L., Almeida, D., Altenschmidt, J., Altman, S., Anadkat, S., et~al.
\newblock Gpt-4 technical report.
\newblock \emph{arXiv preprint arXiv:2303.08774}, 2023.

\bibitem[Anastassiou et~al.(2024)Anastassiou, Chen, Chen, Chen, Chen, Chen, Cong, Deng, Ding, Gao, et~al.]{anastassiou2024seed}
Anastassiou, P., Chen, J., Chen, J., Chen, Y., Chen, Z., Chen, Z., Cong, J., Deng, L., Ding, C., Gao, L., et~al.
\newblock Seed-tts: A family of high-quality versatile speech generation models.
\newblock \emph{arXiv preprint arXiv:2406.02430}, 2024.

\bibitem[Anonymous(2025)]{anonymous2025scaling}
Anonymous.
\newblock Scaling transformers for low-bitrate high-quality speech coding.
\newblock In \emph{The Thirteenth International Conference on Learning Representations}, 2025.
\newblock URL \url{https://openreview.net/forum?id=4YpMrGfldX}.

\bibitem[Barrault et~al.(2023)Barrault, Chung, Meglioli, Dale, Dong, Duppenthaler, Duquenne, Ellis, Elsahar, Haaheim, et~al.]{barrault2023seamless}
Barrault, L., Chung, Y.-A., Meglioli, M.~C., Dale, D., Dong, N., Duppenthaler, M., Duquenne, P.-A., Ellis, B., Elsahar, H., Haaheim, J., et~al.
\newblock Seamless: Multilingual expressive and streaming speech translation.
\newblock \emph{arXiv preprint arXiv:2312.05187}, 2023.

\bibitem[Betker(2023)]{tortoisetts}
Betker, J.
\newblock Better speech synthesis through scaling.
\newblock \emph{CoRR}, abs/2305.07243, 2023.

\bibitem[Borsos et~al.(2023)Borsos, Marinier, Vincent, Kharitonov, Pietquin, Sharifi, Roblek, Teboul, Grangier, Tagliasacchi, et~al.]{borsos2023audiolm}
Borsos, Z., Marinier, R., Vincent, D., Kharitonov, E., Pietquin, O., Sharifi, M., Roblek, D., Teboul, O., Grangier, D., Tagliasacchi, M., et~al.
\newblock Audiolm: a language modeling approach to audio generation.
\newblock \emph{IEEE/ACM transactions on audio, speech, and language processing}, 31:\penalty0 2523--2533, 2023.

\bibitem[Brown et~al.(2020)Brown, Mann, Ryder, Subbiah, Kaplan, Dhariwal, Neelakantan, Shyam, Sastry, Askell, et~al.]{brown2020language}
Brown, T., Mann, B., Ryder, N., Subbiah, M., Kaplan, J.~D., Dhariwal, P., Neelakantan, A., Shyam, P., Sastry, G., Askell, A., et~al.
\newblock Language models are few-shot learners.
\newblock \emph{Advances in neural information processing systems}, 33:\penalty0 1877--1901, 2020.

\bibitem[Chen et~al.(2021)Chen, Chai, Wang, Du, Zhang, Weng, Su, Povey, Trmal, Zhang, et~al.]{chen2021gigaspeech}
Chen, G., Chai, S., Wang, G., Du, J., Zhang, W.-Q., Weng, C., Su, D., Povey, D., Trmal, J., Zhang, J., et~al.
\newblock Gigaspeech: An evolving, multi-domain asr corpus with 10,000 hours of transcribed audio.
\newblock \emph{arXiv preprint arXiv:2106.06909}, 2021.

\bibitem[Chen et~al.(2022)Chen, Wang, Chen, Wu, Liu, Chen, Li, Kanda, Yoshioka, Xiao, et~al.]{chen2022wavlm}
Chen, S., Wang, C., Chen, Z., Wu, Y., Liu, S., Chen, Z., Li, J., Kanda, N., Yoshioka, T., Xiao, X., et~al.
\newblock Wavlm: Large-scale self-supervised pre-training for full stack speech processing.
\newblock \emph{IEEE Journal of Selected Topics in Signal Processing}, 16\penalty0 (6):\penalty0 1505--1518, 2022.

\bibitem[Chen et~al.(2024{\natexlab{a}})Chen, Liu, Zhou, Liu, Tan, Li, Zhao, Qian, and Wei]{chen2024vall}
Chen, S., Liu, S., Zhou, L., Liu, Y., Tan, X., Li, J., Zhao, S., Qian, Y., and Wei, F.
\newblock Vall-e 2: Neural codec language models are human parity zero-shot text to speech synthesizers.
\newblock \emph{arXiv preprint arXiv:2406.05370}, 2024{\natexlab{a}}.

\bibitem[Chen et~al.(2024{\natexlab{b}})Chen, Niu, Ma, Deng, Wang, Zhao, Yu, and Chen]{chen2024f5}
Chen, Y., Niu, Z., Ma, Z., Deng, K., Wang, C., Zhao, J., Yu, K., and Chen, X.
\newblock F5-tts: A fairytaler that fakes fluent and faithful speech with flow matching.
\newblock \emph{arXiv preprint arXiv:2410.06885}, 2024{\natexlab{b}}.

\bibitem[D{\'e}fossez et~al.(2022)D{\'e}fossez, Copet, Synnaeve, and Adi]{defossez2022high}
D{\'e}fossez, A., Copet, J., Synnaeve, G., and Adi, Y.
\newblock High fidelity neural audio compression.
\newblock \emph{arXiv preprint arXiv:2210.13438}, 2022.

\bibitem[D{\'e}fossez et~al.(2024)D{\'e}fossez, Mazar{\'e}, Orsini, Royer, P{\'e}rez, J{\'e}gou, Grave, and Zeghidour]{defossez2024moshi}
D{\'e}fossez, A., Mazar{\'e}, L., Orsini, M., Royer, A., P{\'e}rez, P., J{\'e}gou, H., Grave, E., and Zeghidour, N.
\newblock Moshi: a speech-text foundation model for real-time dialogue.
\newblock \emph{arXiv preprint arXiv:2410.00037}, 2024.

\bibitem[Du et~al.(2024{\natexlab{a}})Du, Chen, Zhang, Hu, Lu, Yang, Hu, Zheng, Gu, Ma, et~al.]{du2024cosyvoice}
Du, Z., Chen, Q., Zhang, S., Hu, K., Lu, H., Yang, Y., Hu, H., Zheng, S., Gu, Y., Ma, Z., et~al.
\newblock Cosyvoice: A scalable multilingual zero-shot text-to-speech synthesizer based on supervised semantic tokens.
\newblock \emph{arXiv preprint arXiv:2407.05407}, 2024{\natexlab{a}}.

\bibitem[Du et~al.(2024{\natexlab{b}})Du, Wang, Chen, Shi, Lv, Zhao, Gao, Yang, Gao, Wang, et~al.]{du2024cosyvoice2}
Du, Z., Wang, Y., Chen, Q., Shi, X., Lv, X., Zhao, T., Gao, Z., Yang, Y., Gao, C., Wang, H., et~al.
\newblock Cosyvoice 2: Scalable streaming speech synthesis with large language models.
\newblock \emph{arXiv preprint arXiv:2412.10117}, 2024{\natexlab{b}}.

\bibitem[Eskimez et~al.(2024)Eskimez, Wang, Thakker, Li, Tsai, Xiao, Yang, Zhu, Tang, Tan, et~al.]{eskimez2024e2}
Eskimez, S.~E., Wang, X., Thakker, M., Li, C., Tsai, C.-H., Xiao, Z., Yang, H., Zhu, Z., Tang, M., Tan, X., et~al.
\newblock E2 tts: Embarrassingly easy fully non-autoregressive zero-shot tts.
\newblock In \emph{2024 IEEE Spoken Language Technology Workshop (SLT)}, pp.\  682--689. IEEE, 2024.

\bibitem[Gao et~al.(2023{\natexlab{a}})Gao, Li, Wang, Luo, Shi, Chen, Li, Zuo, Du, Xiao, et~al.]{gao2023funasr}
Gao, Z., Li, Z., Wang, J., Luo, H., Shi, X., Chen, M., Li, Y., Zuo, L., Du, Z., Xiao, Z., et~al.
\newblock Funasr: A fundamental end-to-end speech recognition toolkit.
\newblock \emph{arXiv preprint arXiv:2305.11013}, 2023{\natexlab{a}}.

\bibitem[Gao et~al.(2023{\natexlab{b}})Gao, Zhang, McLoughlin, and Yan]{gao2023paraformerfastaccurateparallel}
Gao, Z., Zhang, S., McLoughlin, I., and Yan, Z.
\newblock Paraformer: Fast and accurate parallel transformer for non-autoregressive end-to-end speech recognition, 2023{\natexlab{b}}.
\newblock URL \url{https://arxiv.org/abs/2206.08317}.

\bibitem[Guo et~al.(2024)Guo, Liu, Shen, Wu, Xie, Xie, and Xu]{guo2024fireredtts}
Guo, H.-H., Liu, K., Shen, F.-Y., Wu, Y.-C., Xie, F.-L., Xie, K., and Xu, K.-T.
\newblock Fireredtts: A foundation text-to-speech framework for industry-level generative speech applications.
\newblock \emph{arXiv preprint arXiv:2409.03283}, 2024.

\bibitem[Han et~al.(2024)Han, Zhou, Liu, Chen, Meng, Qian, Liu, Zhao, Li, and Wei]{han2024vall}
Han, B., Zhou, L., Liu, S., Chen, S., Meng, L., Qian, Y., Liu, Y., Zhao, S., Li, J., and Wei, F.
\newblock Vall-e r: Robust and efficient zero-shot text-to-speech synthesis via monotonic alignment.
\newblock \emph{arXiv preprint arXiv:2406.07855}, 2024.

\bibitem[He et~al.(2024)He, Shang, Wang, Li, Gu, Hua, Liu, Yang, Li, Shi, et~al.]{he2024emilia}
He, H., Shang, Z., Wang, C., Li, X., Gu, Y., Hua, H., Liu, L., Yang, C., Li, J., Shi, P., et~al.
\newblock Emilia: An extensive, multilingual, and diverse speech dataset for large-scale speech generation.
\newblock In \emph{2024 IEEE Spoken Language Technology Workshop (SLT)}, pp.\  885--890. IEEE, 2024.

\bibitem[Hu et~al.(2021)Hu, Shen, Wallis, Allen-Zhu, Li, Wang, Wang, and Chen]{hu2021lora}
Hu, E.~J., Shen, Y., Wallis, P., Allen-Zhu, Z., Li, Y., Wang, S., Wang, L., and Chen, W.
\newblock Lora: Low-rank adaptation of large language models.
\newblock \emph{arXiv preprint arXiv:2106.09685}, 2021.

\bibitem[Jaech et~al.(2024)Jaech, Kalai, Lerer, Richardson, El-Kishky, Low, Helyar, Madry, Beutel, Carney, et~al.]{jaech2024openai}
Jaech, A., Kalai, A., Lerer, A., Richardson, A., El-Kishky, A., Low, A., Helyar, A., Madry, A., Beutel, A., Carney, A., et~al.
\newblock Openai o1 system card.
\newblock \emph{arXiv preprint arXiv:2412.16720}, 2024.

\bibitem[Ji et~al.(2024)Ji, Jiang, Wang, Chen, Fang, Zuo, Yang, Cheng, Wang, Li, et~al.]{ji2024wavtokenizer}
Ji, S., Jiang, Z., Wang, W., Chen, Y., Fang, M., Zuo, J., Yang, Q., Cheng, X., Wang, Z., Li, R., et~al.
\newblock Wavtokenizer: an efficient acoustic discrete codec tokenizer for audio language modeling.
\newblock \emph{arXiv preprint arXiv:2408.16532}, 2024.

\bibitem[Ji et~al.(2025)Ji, Li, Ye, Wu, Xu, Mo, and Zhang]{ji2025test}
Ji, Y., Li, J., Ye, H., Wu, K., Xu, J., Mo, L., and Zhang, M.
\newblock Test-time computing: from system-1 thinking to system-2 thinking.
\newblock \emph{arXiv preprint arXiv:2501.02497}, 2025.

\bibitem[Kang et~al.(2024)Kang, Yang, Yao, Kuang, Yang, Guo, Lin, and Povey]{kang2024libriheavy}
Kang, W., Yang, X., Yao, Z., Kuang, F., Yang, Y., Guo, L., Lin, L., and Povey, D.
\newblock Libriheavy: a 50,000 hours asr corpus with punctuation casing and context.
\newblock In \emph{ICASSP 2024-2024 IEEE International Conference on Acoustics, Speech and Signal Processing (ICASSP)}, pp.\  10991--10995. IEEE, 2024.

\bibitem[Kaplan et~al.(2020)Kaplan, McCandlish, Henighan, Brown, Chess, Child, Gray, Radford, Wu, and Amodei]{kaplan2020scaling}
Kaplan, J., McCandlish, S., Henighan, T., Brown, T.~B., Chess, B., Child, R., Gray, S., Radford, A., Wu, J., and Amodei, D.
\newblock Scaling laws for neural language models.
\newblock \emph{arXiv preprint arXiv:2001.08361}, 2020.

\bibitem[Kharitonov et~al.(2023)Kharitonov, Vincent, Borsos, Marinier, Girgin, Pietquin, Sharifi, Tagliasacchi, and Zeghidour]{speartts}
Kharitonov, E., Vincent, D., Borsos, Z., Marinier, R., Girgin, S., Pietquin, O., Sharifi, M., Tagliasacchi, M., and Zeghidour, N.
\newblock Speak, read and prompt: High-fidelity text-to-speech with minimal supervision.
\newblock \emph{Trans. Assoc. Comput. Linguistics}, 11:\penalty0 1703--1718, 2023.

\bibitem[Kim et~al.()Kim, Lee, Chung, and Cho]{kimclam}
Kim, J., Lee, K., Chung, S., and Cho, J.
\newblock Clam-tts: Improving neural codec language model for zero-shot text-to-speech.
\newblock In \emph{The Twelfth International Conference on Learning Representations}.

\bibitem[Kong et~al.(2020)Kong, Kim, and Bae]{kong2020hifi}
Kong, J., Kim, J., and Bae, J.
\newblock Hifi-gan: Generative adversarial networks for efficient and high fidelity speech synthesis.
\newblock \emph{Advances in neural information processing systems}, 33:\penalty0 17022--17033, 2020.

\bibitem[Kumar et~al.(2024)Kumar, Seetharaman, Luebs, Kumar, and Kumar]{kumar2024high}
Kumar, R., Seetharaman, P., Luebs, A., Kumar, I., and Kumar, K.
\newblock High-fidelity audio compression with improved rvqgan.
\newblock \emph{Advances in Neural Information Processing Systems}, 36, 2024.

\bibitem[Lajszczak et~al.(2024)Lajszczak, C{\'{a}}mbara, Li, Beyhan, van Korlaar, Yang, Joly, Mart{\'{\i}}n{-}Cortinas, Abbas, Michalski, Moinet, Karlapati, Muszynska, Guo, Putrycz, Gambino, Yoo, Sokolova, and Drugman]{basetts}
Lajszczak, M., C{\'{a}}mbara, G., Li, Y., Beyhan, F., van Korlaar, A., Yang, F., Joly, A., Mart{\'{\i}}n{-}Cortinas, {\'{A}}., Abbas, A., Michalski, A., Moinet, A., Karlapati, S., Muszynska, E., Guo, H., Putrycz, B., Gambino, S.~L., Yoo, K., Sokolova, E., and Drugman, T.
\newblock {BASE} {TTS:} lessons from building a billion-parameter text-to-speech model on 100k hours of data.
\newblock \emph{CoRR}, abs/2402.08093, 2024.

\bibitem[{\L}ajszczak et~al.(2024){\L}ajszczak, C{\'a}mbara, Li, Beyhan, van Korlaar, Yang, Joly, Mart{\'\i}n-Cortinas, Abbas, Michalski, et~al.]{lajszczak2024base}
{\L}ajszczak, M., C{\'a}mbara, G., Li, Y., Beyhan, F., van Korlaar, A., Yang, F., Joly, A., Mart{\'\i}n-Cortinas, {\'A}., Abbas, A., Michalski, A., et~al.
\newblock Base tts: Lessons from building a billion-parameter text-to-speech model on 100k hours of data.
\newblock \emph{arXiv preprint arXiv:2402.08093}, 2024.

\bibitem[Le et~al.(2024)Le, Vyas, Shi, Karrer, Sari, Moritz, Williamson, Manohar, Adi, Mahadeokar, et~al.]{le2024voicebox}
Le, M., Vyas, A., Shi, B., Karrer, B., Sari, L., Moritz, R., Williamson, M., Manohar, V., Adi, Y., Mahadeokar, J., et~al.
\newblock Voicebox: Text-guided multilingual universal speech generation at scale.
\newblock \emph{Advances in neural information processing systems}, 36, 2024.

\bibitem[Li et~al.(2023)Li, Liu, Lam, Wu, Weng, and Meng]{li2023diverse}
Li, X., Liu, S., Lam, M.~W., Wu, Z., Weng, C., and Meng, H.
\newblock Diverse and expressive speech prosody prediction with denoising diffusion probabilistic model.
\newblock \emph{arXiv preprint arXiv:2305.16749}, 2023.

\bibitem[Liu et~al.(2024)Liu, Xu, Yuan, Wu, Wang, and Plumbley]{liu2024semanticodec}
Liu, H., Xu, X., Yuan, Y., Wu, M., Wang, W., and Plumbley, M.~D.
\newblock Semanticodec: An ultra low bitrate semantic audio codec for general sound.
\newblock \emph{arXiv preprint arXiv:2405.00233}, 2024.

\bibitem[Ma et~al.(2024)Ma, Guo, Song, Jiang, Wang, Xue, Xu, Zhao, Zhang, and Xie]{ma2024wenetspeech4tts}
Ma, L., Guo, D., Song, K., Jiang, Y., Wang, S., Xue, L., Xu, W., Zhao, H., Zhang, B., and Xie, L.
\newblock Wenetspeech4tts: A 12,800-hour mandarin tts corpus for large speech generation model benchmark.
\newblock \emph{arXiv preprint arXiv:2406.05763}, 2024.

\bibitem[Ma et~al.(2025)Ma, Tong, Jia, Hu, Su, Zhang, Yang, Li, Jaakkola, Jia, et~al.]{ma2025inference}
Ma, N., Tong, S., Jia, H., Hu, H., Su, Y.-C., Zhang, M., Yang, X., Li, Y., Jaakkola, T., Jia, X., et~al.
\newblock Inference-time scaling for diffusion models beyond scaling denoising steps.
\newblock \emph{arXiv preprint arXiv:2501.09732}, 2025.

\bibitem[Ma et~al.(2023)Ma, Zheng, Ye, Li, Gao, Zhang, and Chen]{ma2023emotion2vec}
Ma, Z., Zheng, Z., Ye, J., Li, J., Gao, Z., Zhang, S., and Chen, X.
\newblock emotion2vec: Self-supervised pre-training for speech emotion representation.
\newblock \emph{arXiv preprint arXiv:2312.15185}, 2023.

\bibitem[Meng et~al.(2024{\natexlab{a}})Meng, Zhou, Liu, Chen, Han, Hu, Liu, Li, Zhao, Wu, Meng, and Wei]{melle}
Meng, L., Zhou, L., Liu, S., Chen, S., Han, B., Hu, S., Liu, Y., Li, J., Zhao, S., Wu, X., Meng, H., and Wei, F.
\newblock Autoregressive speech synthesis without vector quantization.
\newblock \emph{CoRR}, abs/2407.08551, 2024{\natexlab{a}}.

\bibitem[Meng et~al.(2024{\natexlab{b}})Meng, Zhou, Liu, Chen, Han, Hu, Liu, Li, Zhao, Wu, et~al.]{meng2024autoregressive}
Meng, L., Zhou, L., Liu, S., Chen, S., Han, B., Hu, S., Liu, Y., Li, J., Zhao, S., Wu, X., et~al.
\newblock Autoregressive speech synthesis without vector quantization.
\newblock \emph{arXiv preprint arXiv:2407.08551}, 2024{\natexlab{b}}.

\bibitem[Mentzer et~al.(2024)Mentzer, Minnen, Agustsson, and Tschannen]{mentzer2024finite}
Mentzer, F., Minnen, D., Agustsson, E., and Tschannen, M.
\newblock Finite scalar quantization: {VQ}-{VAE} made simple.
\newblock In \emph{The Twelfth International Conference on Learning Representations}, 2024.
\newblock URL \url{https://openreview.net/forum?id=8ishA3LxN8}.

\bibitem[Panayotov et~al.(2015)Panayotov, Chen, Povey, and Khudanpur]{panayotov2015librispeech}
Panayotov, V., Chen, G., Povey, D., and Khudanpur, S.
\newblock Librispeech: an asr corpus based on public domain audio books.
\newblock In \emph{2015 IEEE international conference on acoustics, speech and signal processing (ICASSP)}, pp.\  5206--5210. IEEE, 2015.

\bibitem[Parker et~al.(2024)Parker, Smirnov, Pons, Carr, Zukowski, Evans, and Liu]{parker2024scaling}
Parker, J.~D., Smirnov, A., Pons, J., Carr, C., Zukowski, Z., Evans, Z., and Liu, X.
\newblock Scaling transformers for low-bitrate high-quality speech coding.
\newblock \emph{arXiv preprint arXiv:2411.19842}, 2024.

\bibitem[Pratap et~al.(2020{\natexlab{a}})Pratap, Xu, Sriram, Synnaeve, and Collobert]{Pratap2020MLSAL}
Pratap, V., Xu, Q., Sriram, A., Synnaeve, G., and Collobert, R.
\newblock Mls: A large-scale multilingual dataset for speech research.
\newblock \emph{ArXiv}, abs/2012.03411, 2020{\natexlab{a}}.

\bibitem[Pratap et~al.(2020{\natexlab{b}})Pratap, Xu, Sriram, Synnaeve, and Collobert]{pratap2020mls}
Pratap, V., Xu, Q., Sriram, A., Synnaeve, G., and Collobert, R.
\newblock Mls: A large-scale multilingual dataset for speech research.
\newblock \emph{arXiv preprint arXiv:2012.03411}, 2020{\natexlab{b}}.

\bibitem[Radford et~al.(2019)Radford, Wu, Child, Luan, Amodei, Sutskever, et~al.]{radford2019language}
Radford, A., Wu, J., Child, R., Luan, D., Amodei, D., Sutskever, I., et~al.
\newblock Language models are unsupervised multitask learners.
\newblock \emph{OpenAI blog}, 1\penalty0 (8):\penalty0 9, 2019.

\bibitem[Radford et~al.(2023)Radford, Kim, Xu, Brockman, McLeavey, and Sutskever]{radford2023robust}
Radford, A., Kim, J.~W., Xu, T., Brockman, G., McLeavey, C., and Sutskever, I.
\newblock Robust speech recognition via large-scale weak supervision.
\newblock In \emph{International conference on machine learning}, pp.\  28492--28518. PMLR, 2023.

\bibitem[Reddy et~al.(2021)Reddy, Gopal, and Cutler]{reddy2021dnsmos}
Reddy, C.~K., Gopal, V., and Cutler, R.
\newblock Dnsmos: A non-intrusive perceptual objective speech quality metric to evaluate noise suppressors.
\newblock In \emph{ICASSP 2021-2021 IEEE International Conference on Acoustics, Speech and Signal Processing (ICASSP)}, pp.\  6493--6497. IEEE, 2021.

\bibitem[Ren et~al.(2022)Ren, Hu, Tan, Qin, Zhao, Zhao, and Liu]{ren2022fastspeech2fasthighquality}
Ren, Y., Hu, C., Tan, X., Qin, T., Zhao, S., Zhao, Z., and Liu, T.-Y.
\newblock Fastspeech 2: Fast and high-quality end-to-end text to speech, 2022.
\newblock URL \url{https://arxiv.org/abs/2006.04558}.

\bibitem[Saeki et~al.(2022)Saeki, Xin, Nakata, Koriyama, Takamichi, and Saruwatari]{saeki2022utmos}
Saeki, T., Xin, D., Nakata, W., Koriyama, T., Takamichi, S., and Saruwatari, H.
\newblock Utmos: Utokyo-sarulab system for voicemos challenge 2022.
\newblock \emph{arXiv preprint arXiv:2204.02152}, 2022.

\bibitem[Siuzdak()]{siuzdakvocos}
Siuzdak, H.
\newblock Vocos: Closing the gap between time-domain and fourier-based neural vocoders for high-quality audio synthesis.
\newblock In \emph{The Twelfth International Conference on Learning Representations}.

\bibitem[Snell et~al.(2024)Snell, Lee, Xu, and Kumar]{snell2024scaling}
Snell, C., Lee, J., Xu, K., and Kumar, A.
\newblock Scaling llm test-time compute optimally can be more effective than scaling model parameters.
\newblock \emph{arXiv preprint arXiv:2408.03314}, 2024.

\bibitem[Song et~al.(2024)Song, Chen, Wang, Ma, and Chen]{song2024ella}
Song, Y., Chen, Z., Wang, X., Ma, Z., and Chen, X.
\newblock Ella-v: Stable neural codec language modeling with alignment-guided sequence reordering.
\newblock \emph{arXiv preprint arXiv:2401.07333}, 2024.

\bibitem[Touvron et~al.(2023)Touvron, Lavril, Izacard, Martinet, Lachaux, Lacroix, Rozi{\`e}re, Goyal, Hambro, Azhar, et~al.]{touvron2023Llama}
Touvron, H., Lavril, T., Izacard, G., Martinet, X., Lachaux, M.-A., Lacroix, T., Rozi{\`e}re, B., Goyal, N., Hambro, E., Azhar, F., et~al.
\newblock Llama: Open and efficient foundation language models.
\newblock \emph{arXiv preprint arXiv:2302.13971}, 2023.

\bibitem[Wang et~al.(2023)Wang, Chen, Wu, Zhang, Zhou, Liu, Chen, Liu, Wang, Li, et~al.]{wang2023neural}
Wang, C., Chen, S., Wu, Y., Zhang, Z., Zhou, L., Liu, S., Chen, Z., Liu, Y., Wang, H., Li, J., et~al.
\newblock Neural codec language models are zero-shot text to speech synthesizers.
\newblock \emph{arXiv preprint arXiv:2301.02111}, 2023.

\bibitem[Wang et~al.(2024)Wang, Zhan, Liu, Zeng, Guo, Zheng, Zhang, Zhang, Zhang, and Wu]{wang2024maskgct}
Wang, Y., Zhan, H., Liu, L., Zeng, R., Guo, H., Zheng, J., Zhang, Q., Zhang, X., Zhang, S., and Wu, Z.
\newblock Maskgct: Zero-shot text-to-speech with masked generative codec transformer.
\newblock \emph{arXiv preprint arXiv:2409.00750}, 2024.

\bibitem[Xin et~al.(2024)Xin, Tan, Takamichi, and Saruwatari]{xin2024bigcodec}
Xin, D., Tan, X., Takamichi, S., and Saruwatari, H.
\newblock Bigcodec: Pushing the limits of low-bitrate neural speech codec.
\newblock \emph{arXiv preprint arXiv:2409.05377}, 2024.

\bibitem[Ye et~al.(2024)Ye, Sun, Lei, Lin, Tan, Dai, Kong, Chen, Pan, Liu, et~al.]{ye2024codec}
Ye, Z., Sun, P., Lei, J., Lin, H., Tan, X., Dai, Z., Kong, Q., Chen, J., Pan, J., Liu, Q., et~al.
\newblock Codec does matter: Exploring the semantic shortcoming of codec for audio language model.
\newblock \emph{arXiv preprint arXiv:2408.17175}, 2024.

\bibitem[Zhang et~al.(2024)Zhang, Zhang, Li, Zhou, and Qiu]{zhang2024speechtokenizer}
Zhang, X., Zhang, D., Li, S., Zhou, Y., and Qiu, X.
\newblock Speechtokenizer: Unified speech tokenizer for speech language models.
\newblock In \emph{The Twelfth International Conference on Learning Representations}, 2024.

\bibitem[Zhang et~al.(2023)Zhang, Zhou, Wang, Chen, Wu, Liu, Chen, Liu, Wang, Li, He, Zhao, and Wei]{vallex}
Zhang, Z., Zhou, L., Wang, C., Chen, S., Wu, Y., Liu, S., Chen, Z., Liu, Y., Wang, H., Li, J., He, L., Zhao, S., and Wei, F.
\newblock Speak foreign languages with your own voice: Cross-lingual neural codec language modeling.
\newblock \emph{CoRR}, abs/2303.03926, 2023.

\bibitem[Zhou et~al.(2021)Zhou, Sisman, Liu, and Li]{zhou2021seen}
Zhou, K., Sisman, B., Liu, R., and Li, H.
\newblock Seen and unseen emotional style transfer for voice conversion with a new emotional speech dataset.
\newblock In \emph{ICASSP 2021-2021 IEEE International Conference on Acoustics, Speech and Signal Processing (ICASSP)}, pp.\  920--924. IEEE, 2021.

\bibitem[Zhu et~al.(2024{\natexlab{a}})Zhu, Li, Liu, Ma, and Wang]{zhu2024survey}
Zhu, X., Li, J., Liu, Y., Ma, C., and Wang, W.
\newblock A survey on model compression for large language models.
\newblock \emph{Transactions of the Association for Computational Linguistics}, 12:\penalty0 1556--1577, 2024{\natexlab{a}}.

\bibitem[Zhu et~al.(2024{\natexlab{b}})Zhu, Tian, and Xie]{kalle}
Zhu, X., Tian, W., and Xie, L.
\newblock Autoregressive speech synthesis with next-distribution prediction.
\newblock \emph{arXiv preprint arXiv:2412.16846}, 2024{\natexlab{b}}.

\end{thebibliography}
\bibliographystyle{icml2025}

\newpage
\appendix
\onecolumn







\section{Test set for text understanding ability}
\label{testset}
\subsection{Chinese Evaluation criteria}
 
\begin{CJK*}{UTF8}{gbsn}
 
\begin{table}[ht]
\centering
\caption{Evaluation Criteria for Chinese Test Set}
\label{table:chinese-evaluation-criteria}
\small 
\begin{tabularx}{\textwidth}{|>{\centering\arraybackslash}m{2.5cm}|>{\arraybackslash}X|>{\arraybackslash}X|>{\arraybackslash}X|}
\hline
\textbf{Category} & Score \textbf{1} & Score \textbf{2} & Score \textbf{3} \\
\hline
\textbf{Emotion} & 
No detectable emotion / 无可检测的情感 & 
Emotion present but not convincingly rendered / 存在情感但表达不够令人信服 & 
Correct emotion recognition and appropriate rendering / 正确识别情感并恰当表达 \\
\hline

\textbf{Paralinguistic} & 
No recognition of paralinguistic cues like interjections / 未识别出语调学关键词，如“哎呀”或“嘘” & 
Attempts to render paralinguistic cues but unnatural / 明确意图表达关键词，但表达不自然 & 
Natural rendering of paralinguistic cues with appropriate emphasis / 自然表达语调学关键词，恰当强调 \\
\hline

\textbf{Chinese Poetry} & 
Fails to capture the poetic structure and imagery / 未能捕捉诗歌的结构和意象 & 
Captures some poetic elements but lacks depth / 捕捉了一些诗歌元素但缺乏深度 & 
Accurately captures the poetic structure, imagery, and emotional depth / 准确捕捉诗歌的结构、意象和情感深度 \\
\hline

\textbf{Polyphonic Characters} & 
Incorrect pronunciation and meaning of polyphonic characters / 多音字发音错误，意义不正确 & 
Attempts correct pronunciation but with minor errors / 尝试正确发音但有小错误 & 
Correct pronunciation and meaning of polyphonic characters / 多音字发音和意义正确 \\
\hline

\textbf{Questions} & 
Intonation pattern incorrect, failing to convey questioning tone / 语调模式不正确，未能传达问句的语气 & 
Intonation pattern largely correct but with minor flaws / 语调模式大体正确，但有细微瑕疵 & 
Correct intonation patterns that clearly convey the questioning nature / 语调模式正确，清晰传达问句的性质 \\
\hline

\textbf{Tongue Twisters} & 
Inability to articulate the tongue twister, resulting in errors / 无法清晰表达绕口令，导致错误 & 
Attempts articulation with some errors / 尝试表达绕口令但有部分错误 & 
Clear and accurate articulation of the tongue twister without errors / 清晰准确地表达绕口令，无错误 \\
\hline

\textbf{Rare Characters} & 
Mispronunciation or incorrect interpretation of rare characters / 生僻字发音错误或解释不正确 & 
Attempts correct pronunciation and interpretation with minor mistakes / 尝试正确发音和解释但有小错误 & 
Accurate pronunciation and insightful interpretation of rare characters / 生僻字发音和解释准确 \\
\hline
\end{tabularx}
\end{table}
 

\subsection{Chinese Samples}

\subsubsection{Emotion}
1. 她激动地喊道：“我做到了！真的做到了！”\\
2. 他愤怒地说：“你再这样，我就受不了了！”\\
3. 她悲伤地低声哭泣：“为什么会这样？”\\
4. 他欣喜地笑着说：“这真是太棒了！”\\
5. 她紧张地结结巴巴：“我不知道该怎么办。”\\
6. 他轻松地说道：“没关系，我们可以解决的。”\\
7. 她失望地叹了口气：“我本以为会更好。”\\
8. 他兴奋地跳了起来：“我们赢了！”\\
9. 她害怕地颤抖：“不要靠近我！”\\
10. 他感激地说：“谢谢你，你帮了我大忙。”\\
11. 她羞涩地笑了笑：“我不敢相信。”\\
12. 他绝望地喊道：“一切都结束了吗？”\\
13. 她自豪地说：“这是我的成就。”\\
14. 他焦虑地问：“我们还能挽回吗？”\\
15. 她愉快地唱起歌来：“今天真是美好的一天！”\\
16. 他疲惫地叹息：“我需要休息。”\\
17. 她兴奋地说道：“快看，那里有烟花！”\\
18. 他冷静地回答：“我们需要保持镇定。”\\
19. 她惊讶地说：“这是真的吗？”\\
20. 他满足地微笑：“这一切都很值得。”\\

\subsubsection{Paralinguistic}
1. 哎呀，这雨下得噼里啪啦的，看来今天的郊游计划又泡汤喽！\\
2. 哇塞，烟花在夜空中嗖地一声炸开，瞬间绽放出五彩斑斓的光芒，好美呀！\\
3. 哼，他总是大大咧咧的，走路都咚咚咚地响，一点都不安静呢！\\
4. 嘿哟，这箱子可真沉啊，我费了好大劲才吭哧吭哧地搬起来。\\
5. 咦，这只小猫怎么一直喵喵喵地叫呀，是不是饿了呢？\\
6. 哟呵，你看那只小狗，摇着尾巴汪汪汪地跑过来，好可爱呀！\\
7. 唉，那我就有点好奇了。\\
8. 哇，厨房里传来咕噜咕噜的声音，肯定是妈妈煮的汤快好了，好香啊！\\
9. 哈哈，小朋友们在操场上嘻嘻哈哈地玩耍，笑声一阵接着一阵呢！\\
10. 哇哦，海浪拍打着沙滩，发出哗哗的声音，好像在演奏一首美妙的乐章呢！\\
11. 哎呦，这蚊子嗡嗡嗡地在耳边飞来飞去，烦死我了！\\
12. 哎呀妈呀，这只大鹅伸长了脖子嘎嘎嘎地叫着，气势汹汹地朝我冲过来。\\
13. “嗯，我觉得这主意不错，”她迟疑地说。\\
14. “嘘，小声点，我们不能被发现，”他低语道。\\
15. “额，我不知道该怎么说，”她尴尬地回应。\\
16. “哦，没关系，我可以处理，”她自信地说。\\
17. “啊，天哪！”她惊讶地喊道。\\
18. “呃，好吧，那我们开始吧，”她决定性地说。\\
19. “嘘，露西，嘘……别吵醒你弟弟。”汤姆小声叮嘱，两人轻手轻脚地走过婴儿房。\\
20. “啊，忘了带钥匙了，”她懊恼地说。\\

\subsubsection{Chinese Poetry}
1. 床前明月光，疑是地上霜。举头望明月，低头思故乡。\\
2. 恰同学少年，风华正茂；书生意气，挥斥方遒。指点江山，激扬文字，粪土当年万户侯。曾记否，到中流击水，浪遏飞舟？\\
3. 人生易老天难老，岁岁重阳。今又重阳，战地黄花分外香。\\
4. 雄关漫道真如铁，而今迈步从头越。从头越，苍山如海，残阳如血。\\
5. 红军不怕远征难，万水千山只等闲。\\
6. 踏遍青山人未老，风景这边独好。\\
7. 江山如此多娇，引无数英雄竞折腰。惜秦皇汉武，略输文采；唐宗宋祖，稍逊风骚。一代天骄，成吉思汗，只识弯弓射大雕。俱往矣，数风流人物，还看今朝。\\
8. 天若有情天亦老，人间正道是沧桑。\\
9. 才饮长沙水，又食武昌鱼。万里长江横渡，极目楚天舒。\\
10. 风雨送春归，飞雪迎春到。已是悬崖百丈冰，犹有花枝俏。俏也不争春，只把春来报。待到山花烂漫时，她在丛中笑。\\
11. 朱雀桥边野草花，乌衣巷口夕阳斜。旧时王谢堂前燕，飞入寻常百姓家。\\
12. 劝君莫惜金缕衣，劝君惜取少年时。花开堪折直须折，莫待无花空折枝。\\
13. 红豆生南国，春来发几枝。愿君多采撷，此物最相思。\\
14. 绿蚁新醅酒，红泥小火炉。晚来天欲雪，能饮一杯无？\\
15. 生当作人杰，死亦为鬼雄。至今思项羽，不肯过江东。\\
16. 遥想公瑾当年，小乔初嫁了，雄姿英发。羽扇纶巾，谈笑间樯橹灰飞烟灭。故国神游，多情应笑我，早生华发。人生如梦，一尊还酹江月。\\
17. 多情自古伤离别，更那堪，冷落清秋节。今宵酒醒何处？杨柳岸，晓风残月。此去经年，应是良辰好景虚设。便纵有千种风情，更与何人说？\\
18. 红藕香残玉簟秋，轻解罗裳，独上兰舟。云中谁寄锦书来？雁字回时，月满西楼。花自飘零水自流，一种相思，两处闲愁。此情无计可消除，才下眉头，却上心头。\\
19. 昨夜雨疏风骤，浓睡不消残酒。试问卷帘人，却道海棠依旧。知否，知否？应是绿肥红瘦。\\
20. 帘外雨潺潺，春意阑珊。罗衾不耐五更寒。梦里不知身是客，一晌贪欢。独自莫凭栏，无限江山。别时容易见时难。流水落花春去也，天上人间。\\

\subsubsection{Polyphonic Characters}
1. 这种弹弓的弹力很强。\\
2. 人参苗长得参差不齐，还能让人参观吗？\\
3. 下午要召开工作会议，你去通知一下张会计。\\
4. 这几张茶几，几乎都要散架了。\\
5. 这里有很多畜牧场，养殖了很多牲畜。\\
6. 人要是行，干一行行一行，一行行行行行，行行行干哪行都行，要是不行，干一行不行一行，一行不行行行不行，行行不行，干哪行都不行。\\
7. 他在长跑比赛中行动迅速，长期的训练让他在行动中表现出色，重视每一次长跑的行动细节。\\
8. 他在行业会议上行动自如，行走于各类行动之间，行使政策得以行动实施，令在场行业内人士纷纷称赞他的行动能力。\\
9. 他得到了一份得力的行政支持，得亏了他，得以顺利行动。\\
10. 她乐于助人，有着乐观的工作态度，在乐队中乐声悠扬，乐迷们对她的乐器演奏赞不绝口。\\
11. 他着急地着手准备，着实需要更多时间完成任务，着迷于工作的细节。\\
12. 在好莱坞的片场，她很好学，也演好了每一个角色，赢得了观众的好评。\\
13. 她干劲十足地干了所有使库房干燥的工作。\\
14. 穿上便服，就可以买到便宜的商品，也可也搭乘便车。\\
15. 这几天天天天气不好。\\
16. 来到杨过曾经生活过的地方，小龙女动情的说：“我也想过过过儿过过的生活。”\\
17. 我有一个小本本本来很干净。\\
18. 今天下雨，我骑车差点摔倒，好在我一把把把把住了。\\
19. 校长说校服上除了校徽别别别的，让你们别别别的你非别别的。\\
20. 你去班上数数数数数不好的有多少。\\

\subsubsection{Questions}
1. 今天的会议通知你收到了吗？咱们几点在哪个会议室集合呀？\\
2. 这部新上映的电影口碑据说很不错，你打算去看吗？看完觉得怎么样？\\
3. 你最近工作那么忙，有时间好好休息吗？身体吃得消吗？\\
4. 周末咱们一起去郊外野餐怎么样？你有空吗？\\
5. 你知道明天的天气预报吗？会不会下雨呀？\\
6. 你在大学学的是什么专业？毕业后从事的工作和专业对口吗？\\
7. 这道数学题我怎么都解不出来，你会做吗？能教教我吗？\\
8. 你喜欢什么类型的音乐？是流行、摇滚还是古典呢？\\
9. 你去过国外旅游吗？最喜欢哪个国家？为什么？\\
10. 你觉得我们这次的项目计划可行吗？还有哪些地方需要改进？\\
11. 难道我们遇到一点困难就应该退缩吗？这可不是我们一贯的作风！\\
12. 父母含辛茹苦把我们养大，我们难道不应该好好孝顺他们吗？\\
13. 浪费粮食这种行为，难道不应该受到谴责吗？\\
14. 保护环境是每个人的责任，难道我们可以置之不理吗？\\
15. 他为了集体利益付出了那么多，我们难道不应该感激他吗？\\
16. 老师每天辛苦备课、批改作业，我们难道不应该尊重他们的劳动成果吗？\\
17. 机会摆在面前，我们难道要眼睁睁地看着它溜走吗？\\
18. 努力学习才能有更好的未来，难道这一点还需要怀疑吗？\\
19. 大家都在为了目标拼搏奋斗，我们难道能偷懒懈怠吗？\\
20. 诚实守信是做人的基本准则，难道我们可以随意违背吗？\\

\subsubsection{Tongue Twisters}
1. 老龙恼怒闹老农，老农恼怒闹老龙。农怒龙恼农更怒，龙恼农怒龙怕农。\\
2. 四是四，十是十，十四是十四，四十是四十。莫把四字说成十，休将十字说成四。若要分清四十和十四，经常练说十和四。\\
3. 石狮寺前有四十四个石狮子，寺前树上结了四十四个涩柿子，四十四个石狮子不吃四十四个涩柿子，四十四个涩柿子倒吃四十四个石狮子。\\
4. 粉红墙上画凤凰，凤凰画在粉红墙。红凤凰、粉凤凰，红粉凤凰、花凤凰。\\
5. 哥哥挎筐过宽沟，快过宽沟看怪狗，光看怪狗瓜筐扣，瓜滚筐空哥怪狗。\\
6. 坡上立着一只鹅，坡下就是一条河。宽宽的河，肥肥的鹅，鹅要过河，河要渡鹅，不知是鹅过河，还是河渡鹅？\\
7. 三哥三嫂子，借给我三斗三升酸枣子。等我上山摘了三升三斗酸枣子，再奉还三哥三嫂子这三斗三升酸枣子。\\
8. 墙上一个窗，窗上一支枪，窗下一箩糠。枪落进了糠，糠埋住了枪。窗要糠让枪，糠要枪上墙，墙要枪上窗。互相不退让，糠赶不走枪，枪也上不了窗和墙。\\
9. 蓝教练是女教练，吕教练是男教练，蓝教练不是男教练，吕教练不是女教练。蓝南是男篮主力，吕楠是女篮主力，吕教练在男篮训练蓝南，蓝教练在女篮训练吕楠。\\
10. 任命是任命，人名是人名，任命人名不能错，错了人名错任命。\\
11. 小华和胖娃，一同种庄稼。小华种棉花，胖娃种西瓜。小华的棉花开了花，胖娃的西瓜结了瓜。小华摘棉花，胖娃摘西瓜。棉花、西瓜装一塌，小华、胖娃笑哈哈。\\
12. 黑豆放在黑斗里，黑斗里边放黑豆，黑豆放黑斗，黑斗放黑豆，不知黑豆放黑斗，还是黑斗放黑豆。\\
13. 石小四和史肖石，一同来到阅览室。石小四年十四，史肖石年四十。年十四的石小四爱看诗词，年四十的史肖石爱看报纸。年四十的史肖石发现了好诗词，忙递给年十四的石小四，年十四的石小四见了好报纸，忙递给年四十的史肖石。\\
14. 天上七颗星，地上七块冰，台上七盏灯，树上七只莺，墙上七枚钉。吭唷吭唷拔脱七枚钉。喔嘘喔嘘赶走七只莺。乒乒乓乓踏坏七块冰。一阵风来吹来七盏灯。一片乌云遮掉七颗星。\\
15. 白老八门前栽了八颗白果树，从北边飞来了八个白八哥儿不知在哪住。白老八拿了八个巴达棍儿要打八个白八哥儿，八个白八哥儿飞上了八颗白果树，不知道白老八拿这八个巴达棍儿打着了八个白八哥儿，还是打着了八颗白果树。\\
16. 针蓝线蓝领子蓝，蓝针蓝线蓝领蓝。蓝针蓝线连蓝领，针蓝线蓝领子蓝。\\
17. 白猫满白毛房里一白猫，白猫满白毛。毛白白猫白，白猫白毛毛。\\
18. 炖冻冬瓜冬瓜冻，冻冬瓜，炖冻冬瓜是炖冻冬瓜，不炖冻冬瓜不是炖冻冬瓜。炖冻冬瓜吃炖冻冬瓜，不炖冻冬瓜不吃炖冻冬瓜。\\
19. 补皮裤皮裤破，补皮裤，皮裤不破不补裤。\\
20. 父母的父母扶父母，父母扶父母的父母。父母是父母的父母，父母的父母是父母。\\

\subsubsection{Rare Characters}
1. 嵇康在刑场抚琴，那曲《广陵散》如黄钟大吕，余音袅袅，展现出他不向世俗低头的狷介（juàn jiè）风骨。\\
2. 林黛玉常常顾影自怜，在潇湘馆里，她的情思如缱绻（qiǎn quǎn）的丝线，缠绕着无尽的哀愁。\\
3. 尼采的思想犹如浩瀚星空中的璀璨流星，以其特立独行的哲思，打破了人们习以为常的谫陋（jiǎn lòu）认知。\\
4. 妲己凭借着自己的妖娆妩媚，在商纣王的宫廷中弄权，她的行为可谓是牝鸡司晨（pìn jī sī chén），加速了商朝的灭亡。\\
5. 李白在月下独酌，酒入愁肠，他的才情如奔涌的江水，挥洒出一篇篇脍炙人口的诗篇，尽显其倜傥不羁（tì tǎng bù jī）的风采。\\
6. 杨绛先生一生笔耕不辍，她的文字温润如玉，蕴含着对生活的深刻洞察，她的智慧与豁达令人肃然起敬，堪称一代懿范（yì fàn）。\\
7. 蒲松龄笔下的狐仙鬼怪，或狡黠灵动，或温婉善良，在他构建的奇幻世界里，演绎着人间的悲欢离合，充满了谲诡（jué guǐ）的色彩。\\
8. 阿基米德在浴缸中顿悟浮力原理的那一刻，灵感如电光石火般闪现，他的智慧光芒照亮了科学发展的漫漫长路，成为了后世敬仰的圭臬（guī niè）。\\
9. 李清照在国破家亡之际，写下了许多凄婉动人的词句，她的愁绪如氤氲（yīn yūn）的雾气，弥漫在字里行间，令人为之动容。\\
10. 苏轼一生宦海沉浮，却始终保持着豁达乐观的心态，他在黄州的岁月里，寄情山水，写下了许多脍炙人口的佳作，尽显其旷达超逸（kuàng dá chāo yì）的情怀。\\
11. 王尔德以其华丽而叛逆的文风著称，他的作品中充满了对传统道德的揶揄（yé yú）和对人性的深刻剖析。\\
12. 阮籍常常在山野间纵酒放歌，他的行为举止看似荒诞不经，实则是对当时黑暗社会的一种隐晦的訾议（zǐ yì）。\\
13. 居里夫人在简陋的实验室里，经过无数次的尝试和失败，终于发现了镭元素，她的坚韧和执着成为了科学界的楷模，是当之无愧的巾帼豪杰（jīn guó háo jié）。\\
14. 卡夫卡的作品中充满了荒诞与迷茫，他笔下的人物常常在孤独和绝望中挣扎，展现出一种难以言喻的幽眇（yōu miǎo）情感。\\
15. 屈原在汨罗江畔徘徊，他的心中充满了对国家和人民的忧虑，最终以投江的方式表达了自己的忠贞不渝，他的精神成为了中华民族的亢宗之子（kàng zōng zhī zǐ）。\\
16. 张爱玲的文字犀利而又细腻，她以独特的视角描绘了旧上海的繁华与落寞，她的才情和孤傲令人侧目，是文坛上一颗璀璨的明珠，散发着独特的姱容修态（kuā róng xiū tài）。\\
17. 王阳明在龙场悟道，创立了心学，他的思想如醍醐灌顶（tí hú guàn dǐng），对后世的哲学发展产生了深远的影响。\\
18. 泰戈尔的诗歌充满了对自然和生命的热爱，他的文字如潺潺溪流，流淌着温暖和希望，他的作品具有一种独特的骀荡（dài dàng）之美。\\
19. 杜甫在战乱年代，目睹了百姓的疾苦，他的诗歌如黄钟毁弃（huáng zhōng huǐ qì），发出了对社会不公的强烈控诉，成为了历史的真实写照。\\
20. 列夫·托尔斯泰在他的作品中，深刻地揭示了人性的善恶美丑，他的思想如振聋发聩（zhèn lóng fā kuì）的钟声，唤醒了人们对生活的思考。\\

\end{CJK*}

\newpage
 
\subsection{English Evaluation critieria}
Copy from Base-TTS \cite{lajszczak2024base} for better reading.
\begin{table}[ht]
    \centering
    \caption{English evaluation criteria, copy from Base-TTS \cite{lajszczak2024base}}
    \label{table:emergent-abilities} 
    \begin{tabular}{p{1.8cm}p{5cm}p{5.5cm}} %
        \toprule
        \textbf{Categories} & \textbf{Example sentence} & \textbf{Evaluation criteria} \\
        \midrule
        \textbf{Compound Nouns} & The Beckhams decided to rent a charming stone-built quaint countryside holiday cottage. & 1 = fails to recognise compound nouns \newline 2 = fails to realise the phrasal stress naturally \newline 3 = natural phrasal stress \\ [2ex]
        \textbf{Emotions} & "Oh my gosh! Are we really going to the Maldives? That’s unbelievable!" Jennie squealed, bouncing on her toes with uncontained glee.  &   1 = no audible emotions \newline 2 = emotion present but insufficient \newline 3 = correct emotion recognition and appropriate rendering \\ [2ex]
        \textbf{Foreign Words} & Mr. Henry, renowned for his mise en place, orchestrated a seven-course meal, each dish a pièce de résistance. & 1 = pronounces foreign words with incorrect anglicized pronunciation \newline 2 = applies foreign accent but not entirely correctly \newline 3 = correct rendering in the intended language or accepted anglicized reading  \\ [2ex]
        \textbf{Paralinguistics} & "Shh, Lucy, shhh, we mustn't wake your baby brother," Tom whispered, as they tiptoed past the nursery. & 1 = no recognition of paralinguistic keywords such as "shhh" or "phew" \newline 2 = clear intention to render keywords distinctly, but rendering unnatural \newline 3 = natural rendering, e.g. making speech voiceless on "shhh" and other whispered speech   \\ [2ex]
        \textbf{Punctuations} & She received an odd text from her brother: 'Emergency @ home; call ASAP! Mom \& Dad are worried...\#familymatters.' & 1 = glitches on uncommon punctuations such as \# or \& \newline 2 = no glitch but incorrect rendering \newline 3 = no glitch and correct pausing and verbalization, e.g. @ as "at".  \\ [2ex]
        \textbf{Questions} & But the Brexit question remains: After all the trials and tribulations, will the ministers find the answers in time? & 1 = intonation pattern incorrect \newline 2 = intonation pattern largely correct but with minor flaws \newline 3 = correct intonation  \\ [2ex]
        \textbf{Syntactic Complexities} & The movie that De Moya who was recently awarded the lifetime achievement award starred in 2022 was a box-office hit, despite the mixed reviews. & 1 = failure to parse the syntax correctly \newline 2 = parses the syntax largely correctly but the rendering is not entirely natural \newline 3 = parsing correct and rendering natural \\
        \bottomrule
    \end{tabular}
\end{table}

\newpage
 
\subsection{English Samples}
 
\subsubsection{Questions}
1. You went to the party, even though I explicitly told you not to?

2. There is another aircraft still in the air???

3. Now, seriously, you're saying I am the one to blame?

4. But she clearly doesn't want to?

5. To Hungary and back?

6. You're a copper?

7. What is Data Informed Diplomacy, with all its various manifestations?

8. What's really happening, and is there more than meets the eye?

9. How on earth is this Financial Report organized?

10. Where has Jason Abhisheki moved all the flowers to? 

11. What do we do in this situation, and what are the implications for Jordan's water supply?

12. But the Brexit question remains: After all the trials and tribulations, will the ministers find the answers in time?

13. Sorry, can you restate your name and address please? 

14. Here's the full story for today, would you like to learn more? 

15. Mr. Chairman, your highly anticipated interview with Channel 4 has turned into a catastrophe, hasn't it? 

16. Johnny boy, don't go around acting tough if you can't back it up, right? 

17. Are you in favor of the Latex usage policy or you're just sucking up to leadership?

18. Is it a bird, or is it a plane?

19. Madam, have you tried turning it off and on again? 

20. Were you the one with the hand-held camera or the one with a weird-looking android phone?

\subsubsection{Emotions}

1. Her hands shaking with excitement, Alice Monroe stuttered, "oh..I-I can't believe it! Is this really my acceptance letter to Harvard?" Marco cannot believe it either: "God damn it! How did you pull this off?"

2. A surge of anger flashed across the face of Matthew, as he bellowed, "You have crossed the line this time, and I won't stand for it any longer! Get out!"

3. Gazing at the panoramic view from a mountain in Iceland, Jeff Farahmand sighed deeply, whispering, "This... this is truly the face of the Divine. What more can I ask for?"

4. "You mustn't underestimate how profoundly I've missed your presence," Ito murmured, his eyes glistening with tears as he embraced his long lost sister. "You're finally back, but where do I find our lost years?"

5. "Oh my gosh! Are we really going to the Maldives? That’s unbelievable!" Jennie squealed, bouncing on her toes with obvious glee.

6. "I can confidently declare that this is the most exquisite chocolate cake my taste buds have ever had the pleasure to encounter!" Mo proclaimed, savoring every bite. He could not stop eating! 

7. A proud smile spread across his face as he softly said, "Son, your accomplishments fill my heart with such joy and pride." But then the smile suddenly ceased. Mike’s hearts were pounding like door knocks. His dad’s face now looks like that of the devil himself. 

8. Choking back sobs, Mahmoud whimpered, "I simply can't fathom a life without you by my side. Don't go!"

9. His voice trembled with palpable fear as he stuttered, "There's... there's a stranger at the window. Where the hell are you all waiting for?!"

10. Tears of joy trickled down her cheeks as she yelled, "Graduating as valedictorian... this is a dream come true!"

11. Jane's eyes wide with terror, she screamed, "The brakes aren't working! What do we do now? We're completely trapped!"

12. A profound sense of realization washed over Beal as he whispered, "You've been there for me all along, haven't you? I never truly appreciated you until now."

13. Beth collapsed into his arms, sobbing uncontrollably, "I failed them, I failed them all. They’re all dead! Nothing we can do will ever bring them back. How can I ever live with myself again? How?"

14. His face lit up with pure delight as he exclaimed, "We did it! We won the championship! I knew we could do it together!"

15. Overcome with guilt, Martin hung his head and muttered, "I'm so sorry. I never meant to hurt you like this. Can you ever forgive me?" It was obvious what the answer would be. 

16. The queen danced around the room, eyes twinkling with mischief, "Guess what? I got the lead role in the play! Can you believe it? Well, I can’t."

17. Staring into the distance, the firefighter said with a melancholic smile, "She used to sit right there, you know. I can still hear her laugh if I close my eyes." Outside the window, the rain was pouring down and gushing through every cracks. 

18. The detective’s voice, full of determination and fire, was heard loud and clear in the room, "No one will tell me what I can or cannot do. I'll prove them all wrong! Get me my gun. What are you all looking at me for?"

19. Overwhelmed with confusion and despair, David Darlan cried out, "What do you want from me? Why can't you just tell me what's wrong? Leave me alone!"

20. With a gentle touch and a loving smile, she reassured, "Don't worry, my love. We'll get through this together, just like we always have. I love you."

\subsubsection{Compound Nouns}

1. In the heart of Lagos, there is a public park with a serene duck pond. Nearby, the children's outdoor play area is full of joyful laughter. Nobody knows the darkness descending soon. 

2. At the family reunion, my grandfather, or father-in-law for some, told many tongue-in-cheek jokes. 

3. The physics teacher asked the students to build a new model solar system. Students were told to bring a tape measure and a pair of scissors, to cut the scale-model planet rings.

4. On this fateful day in 1987, the students boarded the little yellow school bus, chattering excitedly about their field trip to the zoo.

5. Hello, we are representatives from Northern Airlines. Please look out from the big second-floor window seat.

6. After years of work, Heisenberg finally published a ground-breaking cutting-edge research paper on quantum physics.

7. Recipe for a delicious breakfast sandwich: avocado, egg, and cheese on a bagel, cooked over a stovetop frying pan.

8. There is nothing more peaceful than a blue water fountain with a wooden greenhouse. Near there, Joseph installed a hard stone birdbath.

9. Prague, Czechia: Good morning, Harari! Here come the big shopping carts and last-minute video game shoppers.

10. My dog knocked over the tea table and all the books scattered across the second living room floor. 

11. The hiking trail up Yahu Mountain provides a spectacular view of the sunrise. Along the path, the wooden signposts with triple-checked trail maps and green distance markers guided us.

12. The fish clock tower was striking again, reminding us of that profound changing of the guard. 

13. Dean Graham sat on the packed wooden park bench, feeding the pigeons while enjoying the pleasant weather. 

14. The Beckhams decided to rent a charming stone-built quaint countryside holiday cottage. 

15. The construction of the new Newtown-council town hall has made huge trouble; rush-hour traffic jam has never been worse. 

16. Owen Farrell has taken England to the Rugby World Cup glory, with a razor-thin-margin victory against New Zealand in France.

17. Scientists at AWS teams are making last-minute pre-launch model preparations.

18. Bad weather in Northern Europe has caused a god-awful flight check-in time of 6 AM, when even the airport food court isn't open. 

19. Jake Park boasts a beautiful hand-built wooden bird feeder.

20. We visited a quaint bed-and-breakfast establishment, complete with lighthouse lamp room.

\subsubsection{Syntactic Complexity}

1. The complex houses married and single soldiers and their families.

2. Time flies like an arrow; fruit flies like a banana.

3. The rat the cat the dog chased killed ate the malt.

4. After the writer the editor the publisher hired fired quit, the company found itself in quite a bind.

5. The old man the boats on the shore were manned by had a long history of seafaring.

6. Anyone who feels that if so many more students whom we haven't actually admitted are sitting in on the course than ones we have that the room had to be changed, then probably auditors will have to be excluded, is likely to agree that the curriculum needs revision.

7. While John, who had been working late every night for a month on his novel, finally took a break to enjoy the fresh air, his neighbor, a painter who often found inspiration in the midnight moon, was just beginning her creative process.

8. In the old village with its winding roads, colorful marketplaces, a sense of history that permeates every brick, and a single traffic light, you'll find peace and simplicity.

9. The chef seasoning the fish tossed it gently.

10. As the sun dipped below the horizon, casting a golden glow over the ocean, Emily, who had spent her life dreaming of distant shores, stood on the deck of the ship, feeling a mixture of anticipation and nostalgia as her adventure began.

11. During the meeting, where Coke executives debated the future of the company, Thomas, a young intern who had discovered a solution, mustered the courage to speak, shifting the direction of the conversation, that preceded his intervention.

12. The movie that De Moya who was recently awarded the lifetime achievement award starred in 2022 was a box-office hit, despite the mixed reviews.

13. In the garden, where the flowers that the gardener who retired last year still bloomed, the children who play there every afternoon find peace and joy.

14. The scientist, Mateusz Gorka, who proposed the theory, which many experts in the field, including those who had initially been skeptical bordering on disbelieving, now support, was nominated for a prestigious award.

15. Although the meal that the chef, who had just returned from a culinary tour of Italy, prepared was delicious, the Greek guests barely noticed.

16. The book that the woman who the man who the child spoke to this morning was reading became a topic of conversation among the friends who had all read it.

17. Despite the fact that the road that led to the Five Villages, which was known for its scenic beauty, was narrow and winding, tourists flocked there throughout the year.

18. CNN journalists tracking the stories behind the officials who served during the tumultuous period when the protests rocked the nation to its core noticed significant inconsistencies in the official reports provided.

19. The musicians who performed the symphony that the composer, whose work had often been overlooked in his lifetime, wrote in his early years received a standing ovation.

20. Cars displayed in these showrooms with ENERGY-EFFICIENT AND GREEN decals prominently featured across the windshield aren't announcing environmentalism; they're virtue signaling.

\subsubsection{Foreign Words}

1. With an ample supply of joie de vivre, Mary danced through the streets of Nice, stopping only to enjoy a nice cafe with a warm croissant.

2. The modern exhibit was a mélange of styles, from German Expressionism to French Impressionism, capturing the Zeitgeist of the time.

3. As a gesture of camaraderie, the Spanish torero invited his rival, Leo the Monster, to a tapas bar, where they enjoyed jamón ibérico and the noche.

4. During Anthony’s wanderlust-filled travels, he discovered the gemütlich atmosphere of many Austrian villages.

5. CloudCorp’s CEO believes in gesamtkunstwerk, like integrating a symphony into a harmonious ensemble.

6. Mr. Henry, renowned for his mise en place, orchestrated a seven-course meal, each dish a pièce de résistance.

7. The fiesta, filled with música, dance, and the warmth of amigos, continued until dawn, embodying the true spirit of a Catalan celebration.

8. At the G20 Summit, leaders discussed rapprochement, trying to step away from the Schadenfreude of political rivalries.

9. After a tiring day, Sarah treated herself to a spa experience, enjoying the sauna and the jacuzzi, and relaxing with a glass of Riesling. 

10. Lasso's novella, rich in allegory and imbued with a sense of ennui, drew from his experiences living in a French château up near the border.

11. The master from Osaka, Japan, dedicated himself to crafting the perfect "nigiri," with "umami" flavors dancing on the palate.

12. Mikhail Gorbachev's Reforms: Perestroika and Glasnost Define a New Era.

13. Lakshmi's yoga practice, centered around the Sanskrit concept of "ahimsa," influenced her approach to life, mirroring the teachings of Mahatma Gandhi.

14. As they strolled through the Grand Bazaar in Istanbul, they were drawn to the beautiful "kilims," the best of Turkic craftsmanship.

15. Inspired by the ancient Chinese philosophy of "Feng Shui," Li rearranged her house to create a "qi" flow throughout.

16. Embracing the Japanese aesthetic of "wabi-sabi," Hokusai's masterpieces were on full display here.

17. During Rio de Janeiro's famous Carnaval do Brasil, the streets pulsated with the rhythms of "samba".

18. The novel's protagonist, guided by the ancient Greek concept of "arete," seeks excellence and virtue, a journey reminiscent of warrior-philosopher-kings.

19. As an aficionado of Scandinavian design, Ole Gunnarsson appreciated the principle of "hygge," evident in his Danish home.

20. These soldiers - they're supposed to practice with a sense of "bushido", the samurai code of honor, but they're behaving like the imperial beasts they are.

\subsubsection{Punctuations}

1. After a moment of silence, Elena Ivanova finally spoke..., —— her words barely audible over the cracking thunder of a torrential downpour.

2. What!?! You're telling me you've never seen a single episode of 'Game of Thrones' before????! (This was not heard by Prof. Johnson, Dr. Lewis, etc.)

3. "Can anyone hear me over there??? Please, we need help!!! NOW!!!!"

4. “The Power of \& and \% in the Digital Age.” won the first prize in this conference.

5. His latest invention (a device meant to assist in everyday chores (something he never seemed to run out of)), was nothing short of brilliant.

6. She read the label and was surprised to find --- that the "natural" ingredients were actually ..... heavily processed.

7. He relayed his conversation with the bartender, saying, "I told him, 'Your 'signature' cocktail is simply a Margarita with a fancy garnish.'"

8. The presently announced laws were announced in 35°N, 80°W. Specific provisions are to be found in §12 and §17.

9. Please ensure you replace [username] and [password] with your actual credentials before logging in, like jA8!fR3\$mQ1.

10. When Maria asked, 'What's happening tonight?' I replied, 'Well, John — who'll be there at 8:00 p.m. — said, "Let's meet at Sarah's place; bring games, snacks, etc., and don't be late!"'

11. "In the case of Johnson v. Smith, the court found that the defendant's actions — e.g., his failure to fulfill the terms of the contract (see sections 4.1, 4.2, and 4.3), etc. — amounted to a breach of trust.“

12. When asked for his thoughts, he simply replied, «I'll be gone in 5 minutes», and left.

13. I saw Gordon listing the ingredients as follows: <tomatoes>, <fresh basil> (or dried, if unavailable - but it's essential), <olive oil>, <garlic>; salt and pepper.

14. She received an odd text from her brother: 'Emergency @ home; call ASAP! Mom \& Dad are worried...\#familymatters.'

15. The sign at the park's entrance stated, 'Please adhere to the following rules: no littering; no pets (except service animals); no loud music after 9 p.m.'

16. “The Art of /Slash/ and $\backslash$backslash$\backslash$” was the best received talk on modern internet lingo.

17. Jeb's email was brief, but to the point: 'Meeting rescheduled for 3 p.m. tomorrow – apologies for any inconvenience. Regards, J.'.

18. The Dead Sea poems contained several annotations, some of which were quite puzzling: [Section unclear]; [Translation disputed]; [Original wording lost].

19. Her travel blog post was filled with enthusiastic descriptions: 'Best trip ever!!!'; 'Amazing people \& culture!'; 'Can't wait to go back...\#wanderlust.'

20. He shouted, 'Everyone, please gather 'round! Here's the plan: 1) Set-up at 9:15 a.m.; 2) Lunch at 12:00 p.m. (please RSVP!); 3) Playing — e.g., games, music, etc. — from 1:15 to 4:45; and 4) Clean-up at 5 p.m.‘

\subsubsection{Paralinguistics}

1. Principal Dickson began, addressing the Parkside assembly: "Ahem, I'd like to talk to you about something real serious."

2. "Aha! Now I understand," said Staff Sgt. Miller, piecing together the evidence. "The culprit left this behind. Phew."

3. "Ouch! That stings," Lilly cried, as her mother carefully applied the antiseptic. "Not beyond salvation, eh?" She dryly asked. 

4. "Shh, Lucy, sshhh, we mustn't wake your baby brother," Tom whispered, as they tiptoed past the nursery.

5. "Hmm, what do you think, is it too high or two low, um... Dr. Carter?" Haim asked, handing over the instrument. 

6. "Uh, well, Lisa," Tarek stuttered, nervously extending the ring he bought for god-knows how much, "mmm..will you marry me?"

7. "Yawn," Robert said, stretching out on the park bench, "this sunshine makes me sleepy."

8. "Oops! I did it again!" little Katie exclaimed, spilling her milk. 

9. "Whoa, can you believe this, Mike?" Susan said, staring at the intruder. "Wow, you're right. These men ain't meanin' well." 

10. James leaned back in his chair, wiped his forehead, and sighed, "Phew, haha, that was a tough meeting. Thanks for being there, Karen."

11. psst. psst. look right here. 

12. "Aha! I've found it, Professor Green," exclaimed Muzi Han, holding up the rare manuscript. "This could change our entire understanding of history."

13. "Ouch, be careful, Henry!" warned his sister, as he climbed the rickety ladder.

14. David whispered to Emily as the lights dimmed in the theater, "Shh, it's starting." 

15. "Hmm, I don't know about this, Jim," Mary said, looking at the folder paper. "It doesn't seem right."

16. "Uh, are you sure about this?" Tim asked nervously, looking at the steep slope before them. "Whoa, it's higher than I thought," he continued, his voice filled with trepidation. "Aha, but look at the view," Emily responded with excitement, "it's worth the climb!"

17. Ta-da! well? What do you think? This is the best right?

18. "Oops, sorry, Dad!" Jack apologized. "Ugh! you again". Dad was impatient. 

19. "Whoa, what a game, Alex!" Chris exclaimed. "I've never seen anything like that final play."

20. "Phew, we made it, Martha," Tom said, collapsing into the seat after the completion of the Manhattan Project.



\end{document}